\documentclass[%
 reprint,
 amsmath,amssymb,
 aps,
]{revtex4-2}

\usepackage{graphicx}% Include figure files
\usepackage{dcolumn}% Align table columns on decimal point
\usepackage{bm}% bold math
\begin{document}
	\preprint{APS/123-QED}
	
\title{Mass-spectra of singly, doubly, and triply bottom baryons}% Force line breaks with \\
	
	\author{Juhi Oudichhya}
	
	% \altaffiliation[Also at ]{}%Lines break automatically or can be forced with \\
	\author{Keval Gandhi}
	
	\author{Ajay Kumar Rai} 
	\affiliation{Department of Physics, Sardar Vallabhbhai National Institute of Technology, Surat, Gujarat-395007, India}
	%\affiliation{
	%\affiliation{
	%\affiliation{%
	\date{\today}% It is always \today, today,
	
	\begin{abstract}
	The present work has been done on the baryons containing one, two, or three bottom quarks. In this article, the ground-state masses of $\Omega_{b}^{-}$, $\Xi_{bb}^{0}$, $\Xi_{bb}^{-}$, $\Omega_{bb}^{-}$, and $\Omega_{bbb}^{*-}$ baryons are calculated within the framework of Regge phenomenology. Further, the values of Regge slopes and Regge intercepts for singly, doubly, and triply bottom baryons are estimated in both the ($J,M^{2}$) and ($n,M^{2}$) planes to calculate the excited-state masses of these baryons. Here our attempt is to assign a possible spin-parity to recently observed some singly bottom baryons and our results could provide useful information for future experimental searches. Our calculated masses are in agreement with the experimental observations where available and close to other theoretical predictions. 
	\end{abstract}

	\maketitle
	
\section{\label{sec:levela} Introduction} 

Investigations of heavy baryons properties have recently received much attention, both experimentally and theoretically. Studies of excited baryonic states are a crucial aspect of hadron spectroscopy and help to shed light on the mechanisms responsible for the dynamics of quarks and baryon formation. An ideal platform to understand the dynamics of Quantum Chromodynamics (QCD) is to study the properties of baryons that contain a heavy quark (charm $(c)$ or bottom $(b)$). Significant experimental progress has been made in the field of singly bottom baryons in recent years. Until three years ago, there were only two excited bottom baryons, $\Lambda_{b}(5912)^{0}$ and $\Lambda_{b}(5920)^{0}$, which were observed by LHCb and CDF collaboration in 2012 \cite{LHCblambda,CDFlambda}. In the past three years however, many excited singly beauty baryons have been observed:
\begin{enumerate}
\item  The lightest charged $\Sigma_{b}^{\pm}$ ($J^{P}=\frac{1}{2}^{+}$) and $\Sigma_{b}^{*\pm}$ ($J^{P}=\frac{3}{2}^{+}$) baryons have been observed by CDF Collaboration in the $\Lambda_{b}^{0}\pi^{\pm}$ spectrum \cite{CDFSigma,CDFSigma*}. Later, in 2018 the LHCb Collaboration reported two new resonances $\Sigma_{b}(6097)^{\pm}$ in the  $\Lambda_{b}^{0}\pi^{\pm}$ channels \cite{sigmab(6097)} and also the new excited bottom baryon $\Xi_{b}(6227)^{-}$ was observed in both $\Lambda_{b}^{0}K^{-}$ and $\Xi_{b}^{0}\pi^{-}$ invariant mass spectrum \cite{cascade(6227)}.
\item  In 2019 the LHCb Collaboration discovered two new states $\Lambda_{b}(6146)^{0}$ and $\Lambda_{b}(6152)^{0}$ in the $\Lambda_{b}\pi^{+}\pi^{-}$ invariant mass distribution \cite{LHCbLambda}. Later, in 2020 the CMS Collaboration confirmed them and further reported an evidence for a broad excess of events in the $\Lambda_{b}\pi^{+}\pi^{-}$ mass spectrum in the region of 6040-6100 MeV \cite{CMSLambda}. Later, the LHCb confirmed this state to be $\Lambda_{b}(6072)^{0}$ \cite{Lambda2}.
\item Recently, in 2020 four extremely narrow excited $\Omega_{b}^{-}$ states such as $\Omega_{b}(6316)^{-}$, $\Omega_{b}(6330)^{-}$, $\Omega_{b}(6340)^{-}$ and $\Omega_{b}(6350)^{-}$ decaying into $\Xi_{b}^{0}K^{-}$ were observed by the LHCb Collaboration \cite{LHCbOmega}.
\end{enumerate}

However, none of the doubly and triply bottom baryons are experimentally observed yet.
In our previous work, we explained why only five states were observed in $\Omega_{c}^{0}$ baryon and why they are so narrow \cite{Juhi}. Since there are exactly five $1P$ excitations with $J^{P} = 1/2^{-}, 3/2^{-}$ for S=1/2 and $J^{P} = 1/2^{-}, 3/2^{-}, 5/2^{-}$ for S=3/2 if the $ss$ diquark remains in its color triplet spin-1 ground state. But recently the LHCb \cite{LHCbOmega} has observed four out of five extremely narrow excited $\Omega_{b}^{-}$ states, leaving fifth to be predicted and observed. The experimentally missing state at LHCb experiment may have $J^{P}=5/2^{-}$, which is most likely due to the degeneracy with the nearby state at 6350 MeV. If this is the case, it may be hidden in the observed peak at 6350 MeV, which though appears consistent with a single resonance, is actually composed of two. In the recent assignmnet, the same $J^{P}=5/2^{-}$ state is predicted as the unseen state, where mass ranges from 6355 MeV to 6380 MeV \cite{Rosnerassgin}.

The latest Particle Data Group (PDG) \cite{PDG} listed all the singly bottom baryons with their masses, decay widths, spin-parity ($J^{P}$) etc. Ground states of these baryons have been observed (with $J^{P}=\frac{1}{2}^{+}, \frac{3}{2}^{+}$) experimentally, except $\Omega_{b}^{*-}$ with $J^{P}=\frac{3}{2}^{+}$. Unlike other bottom baryons, $\Lambda_{b}^{0}$ have a quite number of established states. The $J^{P}$ values of recently observed excited singly bottom baryons are still missing as shown in Table \ref{tab:table1}. It is very crucial to assign the $J^{P}$ values of the hadrons, which can help to probe their properties such as, form factors, decay widths, branching ratios, hyperfine splitting, etc.  The properties of heavy flavor baryons have been studied using various phenomenological as well as theoretical approaches in the last few years. Recently, a new scheme of state classification called $Jls$ mixing coupling is applied to analyze the masses of the heavy baryons $\Omega_{c,b}$, $\Sigma_{c,b}$, and $\Xi^{'}_{c,b}$ and interprets all the newly LHCb reported resonances of the excited $\Omega_{c,b}$ to be the $P$-wave negative-parity baryons \cite{D. Jia2021}. The authors of Refs. \cite{Thakkar2017,Zalak.bb2016,Zalak.bbb2017} have employed a hypercentral constituent quark model for the calculation of excited-state masses of singly, doubly, and triply heavy baryons. 

Ebert \textit{et al.} \cite{Ebert2011} obtained the mass spectra of heavy baryons in the framework of the QCD-motivated relativistic quark model. Another approach quark-diquark model was used to study the properties of singly heavy flavored baryons in which the structure of baryons is considered as the bound states of quark-diquark pairs instead of the usual three equivalent quarks. To this aim, Bethe Salpeter equation including the Yukawa potential between the constituent quarks of baryons was used \cite{Mohammad}. The authors of Ref. \cite{D.Jia} used Regge approach in the heavy quark-diquark picture and reexamine the orbitally excited spectrum of the charmed and bottom baryons. Their computations on spin-dependent mass splitting suggest that the newly observed baryons $\Xi_{b}(6227)^{-}$, $\Sigma_{b}(6097)^{-}$, $\Sigma_{c}(2800)$, and $\Xi_{c}^{'}(2930)$ are all the 1$P$-wave baryons with spin-parity $J^{P} = \frac{3}{2}^{-}$ preferably. They have also shown the mass predictions of the unobserved $\Xi_{b}$ baryon in the $1P$ and $1D$ states, providing useful information for future experiments. The Ref. \cite{Rosnerassgin} predicted the existence of recently observed four narrow excited $\Omega_{b}^{-}$ baryons with negative parity by assuming that these baryonic states were bound states of a $b$-quark and a $P$-wave $ss$ diquark.

\begin{table}%The best place to locate the table environment is directly after its first reference in text
	\caption{\label{tab:table1}
			Masses and $J^{P}$ values of bottom baryons are listed in PDG \cite{PDG}. The status is given as poor(*), fair(**), and likely(***).
	}
	\begin{ruledtabular}
		\begin{tabular}{lllllllll}
			Resonance  & Mass (MeV) & $J^{P}$ & Status \\
			\colrule\noalign{\smallskip}
			$\Lambda_{b}^{0}$ &  5619.60 $\pm$ 0.17 & $\frac{1}{2}^{+}$ & ***\\
			$\Lambda_{b}(5920)^{0}$ & 5920.09 $\pm$ 0.17 & $\frac{3}{2}^{-}$ & ***\\
			$\Lambda_{b}(6070)^{0}$ & 6072.3 $\pm$ 2.9 & $?^{?}$ &  \\
			$\Lambda_{b}(6146)^{0}$ & 6146.2 $\pm$ 0.4 & $\frac{3}{2}^{+}$ & *** \\
			$\Lambda_{b}(6152)^{0}$ &  6152.5 $\pm$ 0.4 & $\frac{5}{2}^{+}$ & ***\\
			$\Sigma_{b}^{+}$ & 5810.56 $\pm$ 0.25 & $\frac{1}{2}^{+}$ & ***\\
				$\Sigma_{b}^{-}$   & 5815.64 $\pm$ 0.27 & $\frac{1}{2}^{+}$ & ***\\
			$\Sigma_{b}^{*+}$ & 5830.32 $\pm$ 0.25 & $\frac{3}{2}^{+}$ & ***\\
			$\Sigma_{b}^{*-}$  & 5834.74 $\pm$ 0.30 & $\frac{3}{2}^{+}$ & ***\\
			$\Sigma_{b}(6097)^{+}$  & 6095.8 $\pm$ 1.7 & $?^{?}$ & ***\\
				$\Sigma_{b}(6097)^{-}$ & 6098.0 $\pm$ 1.8 & $?^{?}$ & ***\\
			$\Xi_{b}^{0}$  & 5791.9 $\pm$ 0.5 & $\frac{1}{2}^{+}$ & ***\\
			$\Xi_{b}^{-}$   & 5797.0 $\pm$ 0.6 & $\frac{1}{2}^{+}$ & ***\\
			$\Xi_{b}^{'}(5935)^{-}$  & 5935.02 $\pm$ 0.05 & $\frac{1}{2}^{+}$ & ***\\
			$\Xi_{b}(5945)^{0}$  & 5952.3 $\pm$ 0.6 & $\frac{3}{2}^{+}$ & ***\\
				$\Xi_{b}(5955)^{-}$  & 5955.33 $\pm$ 0.13 & $\frac{3}{2}^{+}$ & ***\\
			$\Xi_{b}(6227)^{-}$& 6227.9 $\pm$ 0.9 & $?^{?}$ & ***\\
			$\Omega_{b}^{-}$  & 6046.1 $\pm$ 1.7 & $\frac{1}{2}^{+}$ & ***\\
			$\Omega_{b}(6316)^{-}$ & 6315.6 $\pm$ 0.6 & $?^{?}$ & *\\
			$\Omega_{b}(6330)^{-}$ & 6330.3 $\pm$ 0.6 & $?^{?}$ & *\\
			$\Omega_{b}(6340)^{-}$ & 6339.7 $\pm$ 0.6 & $?^{?}$ & *\\
			$\Omega_{b}(6350)^{-}$  & 6349.8 $\pm$ 0.6 & $?^{?}$ & *\\
		\end{tabular}
	\end{ruledtabular}
\end{table}

\begin{table*}%The best place to locate the table environment is directly after its first reference in text
	\caption{\label{tab:table2}
		Ground state ($J^{P}$=$\frac{1}{2}^{+}$ and $\frac{3}{2}^{+}$) masses of singly, doubly, and triply bottom baryons (in GeV).
	}
	\begin{ruledtabular}
		
		\begin{tabular}{lcccccccccccccccc}
			&\multicolumn{2}{c}{$\Omega_{b}^{-}$} &\multicolumn{2}{c}{$\Xi_{bb}^{0}$/$\Xi_{bb}^{-}$}&\multicolumn{2}{c}{$\Omega_{bb}^{-}$}&\multicolumn{1}{c}{$\Omega_{bbb}^{*-}$} \\
			\colrule\noalign{\smallskip}	  
			$J^{P}$	&\multicolumn{1}{c}{$\frac{1}{2}^{+}$} &\multicolumn{1}{c}{$\frac{3}{2}^{+}$}&\multicolumn{1}{c}{$\frac{1}{2}^{+}$} 	&\multicolumn{1}{c}{$\frac{3}{2}^{+}$} &\multicolumn{1}{c}{$\frac{1}{2}^{+}$}&\multicolumn{1}{c}{$\frac{3}{2}^{+}$} &\multicolumn{1}{c}{$\frac{3}{2}^{+}$}\\
			\colrule\noalign{\smallskip}
			Present &6.054 &6.074 &10.225/10.230 &10.330/10.333 &10.350 &10.449 &14.822  \\
			PDG \cite{PDG} &6.046\\
			Ref. \cite{Wei.2017} &6.048 &6.069 &10.199 &10.316 &10.320 &10.431 &14.788\\
			Ref. \cite{Ebert2005.2.8,Ebert2011} &6.064 &6.088 &10.202 &10.237 &10.359 &10.389\\
			Ref. \cite{Z. Wang 2010.11}&6.110 &6.170 &10.170 &10.220 &10.320 &10.380 &14.830\\
			Ref. \cite{Thakkar2017,Zalak.bb2016,Zalak.bbb2017}&6.048 &6.086&10.321 &10.335 &10.446 &10.467 &14.496\\
			Ref. \cite{Yoshida2015} &6.076 &6.094 &10.314 &10.339 &10.447 &10.467\\
			Ref. \cite{Yamaguchi2015} &6.081 &6.102\\ 
			Ref. \cite{Robert2008} &6.081 &6.102 &10.340 &10.367 &10.454 &10.486 &14.834\\
			Ref. \cite{A.P2008} & &6.102 &10.130 &10.144 &10.422 &10.432 &14.569\\
			Ref. \cite{Roncaglia1995} &6.060 &6.090&10.340 &10.370 &10.370 &10.400\\
			Ref. \cite{D.-H. He2004}& & &10.272 &10.337 &10.369 &10.429\\
			Ref. \cite{Z. Ghalenovi2011}&5.967 &6.096&10.339 &10.468 &10.478 &10.607 &15.118\\
			Ref. \cite{R. Dhir2013}&6.135 &6.142 &10.440 &10.451 &10.620 &10.628 &15.129\\ 
			Ref. \cite{S. P. Tong2000}& & &10.300 &10.340 &10.340 &10.380\\
			Ref. \cite{Z.Brown2014}&6.056 &6.085 &10.143 &10.178 &10.273 &10.308 &14.366\\
			Ref. \cite{S.Agaev2017}&6.024 &6.084\\
			Ref. \cite{J. D. Bjorken1985}& &6.035 & &10.250 & &10.395 &14.760\\
			Ref. \cite{Ghalenovi2014}&5.903 &5.986 &10.334 &10.431 &10.397 &10.495 &15.023\\
			Ref. \cite{Valcarce2008} &6.056 &6.079 &10.189 &10.218 &10.293 &10.321\\
		
			Ref. \cite{Albertus2007} & & &10.197 &10.236 &10.260 &10.297\\
		\end{tabular}
	\end{ruledtabular}
\end{table*}

\begin{table*}%The best place to locate the table environment is directly after its first reference in text
	\caption{\label{tab:table3}
		Regge slopes of ground state $\frac{1}{2}^{+}$ and $\frac{3}{2}^{+}$ trajectories of singly, doubly, and triply bottom baryons (in GeV$^{-2}$) .
	}
	\begin{ruledtabular}
		
		\begin{tabular}{ccccccccccccccccccccc}

			&	$\Lambda_{b}$ $(nnb)$ &$\Sigma_{b}$ $(nnb)$&$\Xi_{b}$ $(nsb)$ &$\Xi_{b}^{'}$ $(nsb)$ &
			$\Omega_{b}$ $(ssb)$ &	$\Xi_{bb}$ $(nbb)$ &$\Omega_{bb}$ $(sbb)$ &$\Omega_{bbb}$ $(bbb)$ \\
			\noalign{\smallskip}\hline\noalign{\smallskip}
			$\alpha^{'}$ &0.2852&0.2852 &0.2795 &0.2795 &0.2740 &0.1792 &0.1656 &-\\
			$\alpha^{'*}$ &-&0.2906 &0.2846 &-&0.2789 &0.1695 &0.1688 &0.1216 \\
		\end{tabular}
	\end{ruledtabular}
\end{table*}
Wei \textit{et al}. have employed the Regge phenomenology and expressed the masses of an unobserved ground state doubly and triply bottom baryons as a function of masses of the well established light baryons and singly bottom baryons. The values of Regge slopes and Regge intercepts are calculated to estimate the masses of the orbitally excited singly, doubly, and triply bottom baryons \cite{Wei.2017}. In the present, we use the same approach of Regge phenomenology with the assumption of linear Regge trajectories. We extract relations between the intercept, slope ratios, and baryon masses in both the $(J,M^{2})$ and $(n,M^{2})$ planes. Using these relations, we derived the expressions to calculate the ground state masses of $\Omega_{b}^{-}$, $\Xi_{bb}^{0}$, $\Xi_{bb}^{-}$, $\Omega_{bb}^{-}$, and $\Omega_{bbb}^{*-}$ baryons. The Regge slopes and Regge intercepts of the $\frac{1}{2}^{+}$ and $\frac{3}{2}^{+}$ trajectories of singly, doubly, and triply bottom baryons are extracted and the masses of the baryon states lying on the $\frac{1}{2}^{+}$ and $\frac{3}{2}^{+}$ trajectories are estimated in both the $(J,M^{2})$ and $(n,M^{2})$ planes. 
Further we extend this scheme and try to calculate the remaining  states other than natural and unnatural parity states in the $(J,M^{2})$ plane. We try to assign a possible spin parity to recently observed excited bottom baryons. This is the main motivation of this paper. 

The remainder of this paper is organized as follows. In Sec. \ref{sec:level1} after briefing of Regge theory, the ground state masses of unobserved bottom baryons $\Omega_{b}^{-}$, $\Xi_{bb}^{0}$, $\Xi_{bb}^{-}$, $\Omega_{bb}^{-}$, and $\Omega_{bbb}^{*-}$ were extracted. After that we estimate the Regge slopes for singly, doubly, and triply bottom baryons for  $\frac{1}{2}^{+}$ and $\frac{3}{2}^{+}$ trajectories, and then the orbitally excited state masses are calculated in the ($J,M^{2}$) plane. In addition, we calculate the radial and orbital excited state masses of these baryons in the ($n,M^{2}$) plane by estimating the Regge slopes and intercepts of particular Regge lines. A discussion of our results is given in Sec \ref{sec:level2}. We concluded our study in Sec IV.

\section{\label{sec:level1}Theoretical Framework}

The plots of Regge trajectories of hadrons in the $(J,M^{2})$ plane are usually called Chew-frautschi plots \cite{Chew1961}. Also, hadrons lying on the same Regge line possess the same internal quantum numbers. The most general form of linear Regge trajectories can be expressed as \cite{Juhi,Wei2008},
\begin{equation}
	\label{eq:1}
	J = \alpha(M) = a(0)+\alpha^{'} M^2 ,
\end{equation}
where $a(0)$ and $\alpha^{'}$ represent the intercept and slope of the trajectory respectively. These parameters for different quark constituents of a baryon multiplet can be related by two relations \cite{Wei2008,Add1,Add2,Add3,Add4},
\begin{equation}
	\label{eq:2}
	a_{iiq}(0) + a_{jjq}(0) = 2a_{ijq}(0) ,	
\end{equation}

\begin{equation}
	\label{eq:3}
	\dfrac{1}{{\alpha^{'}}_{iiq}} + \dfrac{1}{{\alpha^{'}}_{jjq}} = \dfrac{2}{{\alpha^{'}}_{ijq}} ,	
\end{equation}\\
where $i, j, q$ represent quark flavors, $m_{i}<m_{j}$, and $q$ denotes an arbitrary light or heavy quark. Using Eqs. (\ref{eq:1}) and (\ref{eq:2}) we obtain,
\begin{equation}
	\label{eq:4}
	\alpha^{'}_{iiq}M^{2}_{iiq}+\alpha^{'}_{jjq}M^{2}_{jjq}=2\alpha^{'}_{ijq}M^{2}_{ijq} .
\end{equation}

After combining the relations (\ref{eq:3}) and (\ref{eq:4}) we obtain two pairs of solutions as,
\begin{widetext}
	\begin{equation}
		\label{eq:5}
		\dfrac{\alpha^{'}_{jjq}}{\alpha^{'}_{iiq}}=\dfrac{1}{2M^{2}_{jjq}}\times[(4M^{2}_{ijq}-M^{2}_{iiq}-M^{2}_{jjq})
		\pm\sqrt{{{(4M^{2}_{ijq}-M^{2}_{iiq}-M^{2}_{jjq}})^2}-4M^{2}_{iiq}M^{2}_{jjq}}],
	\end{equation}
		and, 
	\begin{equation}
		\label{eq:6}	
		\dfrac{\alpha^{'}_{ijq}}{\alpha^{'}_{iiq}}=\dfrac{1}{4M^{2}_{ijq}}\times[(4M^{2}_{ijq}+M^{2}_{iiq}-M^{2}_{jjq})
		\pm\sqrt{{{(4M^{2}_{ijq}-M^{2}_{iiq}-M^{2}_{jjq}})^2}-4M^{2}_{iiq}M^{2}_{jjq}}].
	\end{equation}
\end{widetext}
This is the important relation we obtained between slope ratios and baryon masses. Many theoretical studies show that the slopes of Regge trajectories decrease with increasing quark masses \cite{Add1,Add2,Zang2007,Brisudova2000,J.L.1986,A.B.1982}. For $m_{j}>m_{i}$, one can write $\alpha^{'}_{jjq}/\alpha^{'}_{iiq}<1$. Consequently from Eq. (\ref{eq:5}), we can say that,
\begin{equation}
	\begin{split}
		\label{eq:7}
	\dfrac{1}{2M^{2}_{jjq}}\times[(4M^{2}_{ijq}-M^{2}_{iiq}-M^{2}_{jjq})\\
		+\sqrt{{{(4M^{2}_{ijq}-M^{2}_{iiq}-M^{2}_{jjq}})^2}-4M^{2}_{iiq}M^{2}_{jjq}}] < 1 ,
	\end{split}
\end{equation}
the above relation gives the inequality equation which is expressed as,
\begin{equation}
	\label{eq:8}
	2M^{2}_{ijq}<M^{2}_{iiq}+M^{2}_{jjq} .
\end{equation}

Now to estimate the deviation of relation (\ref{eq:8}), we introduce a parameter called $\delta$ which is used to replace the sign of inequality with the equal sign. For baryons, it is denoted by $\delta^{b}_{ij,q}$ and given by,
\begin{equation}
	\label{eq:9}
	\delta^{b}_{ij,q} = M^{2}_{iiq}+M^{2}_{jjq}-2M^{2}_{ijq} ,
\end{equation}
here also $i, j$, and $q$ represent the arbitrary light or heavy quarks.
Now from Eqs. (\ref{eq:2}) and (\ref{eq:3}) we can write,
\begin{equation}
	\label{eq:10}
	a_{iiq}(0)-a_{ijq}(0)=a_{ijq}(0)-a_{jjq}(0),
\end{equation}

\begin{equation}
	\label{eq:11}
	\frac{1}{\alpha_{iiq}^{'}}-\frac{1}{\alpha_{ijq}^{'}}=	\frac{1}{\alpha_{ijq}^{'}}-\frac{1}{\alpha_{jjq}^{'}} ,
\end{equation}
based on these equations we introduce two parameters,
\begin{equation}
	\label{eq:12}
	\lambda_{x}=a_{nnn}(0)-a_{nnx}(0)  ,  \gamma_{x}=\frac{1}{\alpha^{'}_{nnx}}-\frac{1}{\alpha^{'}_{nnn}} ;
\end{equation}
where $n$ represents light nonstrange quark ($u$ or $d$) and $x$ denotes $i, j$, or $q$. From Eqs. (\ref{eq:10})-(\ref{eq:12})we have,

\begin{equation}
	\label{eq:13}
	\begin{split}
		a_{ijq}(0) = a_{nnn}(0)-\lambda_{i}-\lambda_{j}-\lambda_{q} ,\\ \frac{1}{\alpha_{ijq}^{'}} = \frac{1}{\alpha_{nnn}^{'}}+\gamma_{i}+\gamma_{j}+\gamma_{q}.
	\end{split}
\end{equation}

Since in baryon multiplets for $nnn$ and $ijq$ states, we can write from Eq. (\ref{eq:4}),
\begin{equation}
	\label{eq:14}
	\begin{split}
		J = a_{nnn}(0) + \alpha_{nnn}^{'}M^{2}_{nnn} ,\\
		J = a_{ijq}(0) + \alpha^{'}_{ijq}M^{2}_{ijq} ,
	\end{split}
\end{equation}

solving Eqs. (\ref{eq:13}) and (\ref{eq:14}) we have,
\begin{equation}
	\label{eq:15}
	M^{2}_{ijq} = (\alpha^{'}_{nnn}M^{2}_{nnn} + \lambda_{i} + \lambda_{j} + \lambda_{q})\left(\frac{1}{\alpha^{'}_{nnn}}+\gamma_{i}+\gamma_{j}+\gamma_{q}\right).
\end{equation}

Combining relations (\ref{eq:9}) and (\ref{eq:15}) we can prove that,
\begin{eqnarray}
	\label{eq:16}
	\nonumber
	\delta^{b}_{ij,q} &=& M^{2}_{iiq}+M^{2}_{jjq}-2M^{2}_{ijq}\\ 
	&=& 2(\lambda_{i}-\lambda_{j})(\gamma_{i}-\gamma_{j}) ,
\end{eqnarray}
which says that $\delta^{b}_{ij,q}$ is independent of quark flavor $q$.

\begin{table}%The best place to locate the table environment is directly after its first reference in text
	\caption{\label{tab:table4}
		Masses of excited states of $\Lambda_{b}^{0}$ baryon in the $(J,M^{2})$ plane for natural parity states. The numbers in the boldface are the experimental values taken as the input \cite{PDG} (in GeV).
	}
	\begin{ruledtabular}
		
		\begin{tabular}{llllllllll}
			\textit{$N^{2S+1}L_{J}$}& Present & PDG \cite{PDG} & \cite{Thakkar2017} & \cite{Ebert2011} & \cite{Wei.2017} & \cite{Ebert2005.2.8} \\	
			\colrule\noalign{\smallskip}
			$1^{2}S_{\frac{1}{2}}$ & \textbf{5.620} & 5.620 & 5.621 & 5.620 &5.619 &5.622 \\
			$1^{2}P_{\frac{3}{2}}$ & 5.924 & 5.920 & 5.988 &5.942 &5.913 & 5.947 \\
			$1^{2}D_{\frac{5}{2}}$ & 6.213 &6.153 &6.213 &6.196 &6.193 &6.197\\
			$1^{2}F_{\frac{7}{2}}$ & 6.489 & &6.432 &6.411 &6.461 &6.405\\
			$1^{2}G_{\frac{9}{2}}$ & 6.754 & & &6.599 &6.718\\
			$1^{2}H_{\frac{11}{2}}$& 7.009\\

		\end{tabular}
	\end{ruledtabular}
\end{table}
\begin{table}%The best place to locate the table environment is directly after its first reference in text
	\caption{\label{tab:table5}
		Masses of excited states of $\Sigma_{b}^{\pm}$ baryon in the $(J,M^{2})$ plane for natural and unnatural parity states. The numbers in the boldface are the experimental values taken as the input \cite{PDG} (in GeV).
	}
	\begin{ruledtabular}
		
		\begin{tabular}{lllllllllllllll}
			\textit{$N^{2S+1}L_{J}$}&\multicolumn{2}{c}{Present}& & Others	\\
			\colrule\noalign{\smallskip}
			&$\Sigma_{b}^{+}$ &$\Sigma_{b}^{-}$ & PDG \cite{PDG} &\cite{Ebert2011} & \cite{Wei.2017} & \cite{Ebert2005.2.8} & \cite{Yoshida2015} \\	
			\colrule\noalign{\smallskip}
			$1^{2}S_{\frac{1}{2}}$ & \textbf{5.811} &\textbf{5.816 }& &5.808 &5.813 &5.805 &5.823 \\
			$1^{2}P_{\frac{3}{2}}$ & 6.105 &6.110 & 6.096 ($\Sigma_{b}^{+}$)&6.096 &6.098 &6.076 &6.132\\
			& && 6.098 ($\Sigma_{b}^{-}$) \\
			$1^{2}D_{\frac{5}{2}}$ & 6.386 &6.390 & &6.284 &6.369 &6.248 &6.397 \\
			$1^{2}F_{\frac{7}{2}}$ & 6.655 &6.659 & &6.500 &6.630  \\
			$1^{2}G_{\frac{9}{2}}$ & 6.913 &6.917 & &6.687 &6.881 \\
			$1^{2}H_{\frac{11}{2}}$& 7.162 &7.166 \\
			\noalign{\smallskip}
			$1^{4}S_{\frac{3}{2}}$ & 5.830 &5.835 &&5.834 &5.834 &5.834 &5.845 \\
			$1^{4}P_{\frac{5}{2}}$ & 6.118 &6.123 &&6.084 &6.117 &6.083 &6.144 \\
			$1^{4}D_{\frac{7}{2}}$ & 6.393 &6.398&&6.260 &6.388 &6.262 \\
			$1^{4}F_{\frac{9}{2}}$ & 6.657 &6.661 &&6.459 &6.648 \\
			$1^{4}G_{\frac{11}{2}}$& 6.910 &6.914 &&6.635 &6.898 \\
			$1^{4}H_{\frac{13}{2}}$& 7.155 &7.158 \\	
			
		\end{tabular}
	\end{ruledtabular}
	
\end{table}

\begin{table}%The best place to locate the table environment is directly after its first reference in text
	\caption{\label{tab:table6}
		Masses of excited states of $\Xi_{b}^{0,-}$ baryon in the $(J,M^{2})$ plane for natural and unnatural parity states. The numbers in the boldface are the experimental values taken as the input \cite{PDG}  (in GeV).
	}
	\begin{ruledtabular}
		
		\begin{tabular}{lllllllllllllll}
			\textit{$N^{2S+1}L_{J}$}&\multicolumn{2}{c}{Present}& & Others	\\
			\colrule\noalign{\smallskip}
			&$\Xi_{b}^{0}$ &$\Xi_{b}^{-}$ &\cite{Ebert2011} & \cite{Wei.2017} & \cite{Ebert2005.2.8} & \cite{B. Chen2015} \\	
			\colrule\noalign{\smallskip}
			$1^{2}S_{\frac{1}{2}}$ & \textbf{5.792 }&\textbf{5.797} &5.803 &5.793 &5.812 &5.801\\
			$1^{2}P_{\frac{3}{2}}$ & 6.093 &6.098 &6.130 &6.080 &6.130 &6.106\\
			$1^{2}D_{\frac{5}{2}}$ & 6.380 &6.385 &6.373 &6.354 &6.365 &6.349 \\
			$1^{2}F_{\frac{7}{2}}$ & 6.654 &6.659 &6.581 &6.616 &6.558 &6.559 \\
			$1^{2}G_{\frac{9}{2}}$ & 6.918 &6.922 &6.762 &6.869 & &6.747\\
			$1^{2}H_{\frac{11}{2}}$& 7.712 &7.176 \\
			\noalign{\smallskip}
			$1^{4}S_{\frac{3}{2}}$ & \textbf{5.952} &\textbf{5.955} &5.963 &5.952 &5.963\\
			$1^{4}P_{\frac{5}{2}}$ & 6.240 &6.243 &6.226 &6.232 &6.218\\ 
			$1^{4}D_{\frac{7}{2}}$ & 6.516 &6.518 &6.414 &6.499 &6.390\\
			$1^{4}F_{\frac{9}{2}}$ & 6.780 &6.782 &6.610 &6.756 \\
			$1^{4}G_{\frac{11}{2}}$& 7.034 &7.036 &6.782 &7.003 \\
			$1^{4}H_{\frac{13}{2}}$& 7.279 &7.281 \\	
		\end{tabular}
	\end{ruledtabular}
\end{table}

\begin{table}%The best place to locate the table environment is directly after its first reference in text
	\caption{\label{tab:table7}
		Masses of excited states of $\Xi_{b}^{'-}$ baryon in the $(J,M^{2})$ plane for natural parity states. The numbers in the boldface are the experimental values taken as the input \cite{PDG} (in GeV).
	}
	\begin{ruledtabular}
		\begin{tabular}{llllllllll}
			\textit{$N^{2S+1}L_{J}$} & Present &PDG \cite{PDG}  & \cite{Ebert2011} & \cite{Wei.2017} & \cite{Ebert2005.2.8} & \cite{Robert2008}\\	
			\colrule\noalign{\smallskip}
			$1^{2}S_{\frac{1}{2}}$ & \textbf{5.935} &5.935 &5.936 &5.935 &5.937 &5.970 \\
			$1^{2}P_{\frac{3}{2}}$ & 6.229 &6.227 &6.234 &6.215 &6.212 &6.190\\
			$1^{2}D_{\frac{5}{2}}$ & 6.510 & &6.432 &6.486 &6.377 &6.393\\
			$1^{2}F_{\frac{7}{2}}$ & 6.779 && 6.641 &6.745\\
			$1^{2}G_{\frac{9}{2}}$ & 7.038 & &6.821 &6.994\\
			$1^{2}H_{\frac{11}{2}}$& 7.288\\
		\end{tabular}
	\end{ruledtabular}
\end{table}

\begin{table}%The best place to locate the table environment is directly after its first reference in text
	\caption{\label{tab:table8}
		Masses of excited states of $\Omega_{b}^{-}$ baryon in the $(J,M^{2})$ plane for natural and unnatural parity states (in GeV).
	}
	\begin{ruledtabular}
		
		\begin{tabular}{lllllllllll}
			\textit{$N^{2S+1}L_{J}$} & Present & PDG \cite{PDG} &\cite{Thakkar2017} &\cite{Ebert2011} &\cite{Wei.2017} &\cite{Yoshida2015}\\	
			\colrule\noalign{\smallskip}
			$1^{2}S_{\frac{1}{2}}$ & 6.054 &6.046 &6.048 &6.064 &6.048 &6.076\\
			$1^{2}P_{\frac{3}{2}}$ & 6.348 &6.340 &6.328 &6.340 &6.325 &6.336 \\
			$1^{2}D_{\frac{5}{2}}$ & 6.629 & &6.567 &6.529 &6.590 &6.561\\
			$1^{2}F_{\frac{7}{2}}$ & 6.899 & &6.800 &6.736 &6.844\\
			$1^{2}G_{\frac{9}{2}}$ & 7.159 & & &6.915 &7.090\\
			$1^{2}H_{\frac{11}{2}}$& 7.409 \\
			\noalign{\smallskip}
			$1^{4}S_{\frac{3}{2}}$ & 6.074 & &6.086 &6.088 &6.069 &6.094 \\
			$1^{4}P_{\frac{5}{2}}$ & 6.362 & &6.320 &6.334 &6.345 &6.345 \\ 
			$1^{4}D_{\frac{7}{2}}$ & 6.638 & &6.553 &6.517 &6.609\\
			$1^{4}F_{\frac{9}{2}}$ & 6.903 & &6.780 &6.713 &6.863 \\
			$1^{4}G_{\frac{11}{2}}$& 7.158 & & &6.884 &7.108 \\
			$1^{4}H_{\frac{13}{2}}$& 7.404 \\	
		\end{tabular}
	\end{ruledtabular}
\end{table}

\begin{table}%The best place to locate the table environment is directly after its first reference in text
	\caption{\label{tab:table9}
		Masses of excited states of $\Xi_{bb}^{0,-}$ baryon in the $(J,M^{2})$ plane for natural and unnatural parity states  (in GeV). 
	}
	\begin{ruledtabular}
		
		\begin{tabular}{llllllllllll}
			\textit{$N^{2S+1}L_{J}$}&\multicolumn{2}{c}{Present}& & Others	\\
			\colrule\noalign{\smallskip}
			&$\Xi_{bb}^{0}$ &$\Xi_{bb}^{-}$ & \cite{Wei.2017} & \cite{Robert2008}& \cite{B. Eakins2012} & \cite{Yoshida2015} \\	
			\colrule\noalign{\smallskip}
			$1^{2}S_{\frac{1}{2}}$ & 10.225 &10.230 &10.199 &10.340 &10.322 &10.314 \\
			$1^{2}P_{\frac{3}{2}}$ & 10.494 &10.499 &10.474 &10.495 &10.692 &10.476\\
			$1^{2}D_{\frac{5}{2}}$ & 10.757 &10.761 &10.742 &10.676 &11.002 &10.592\\
			$1^{2}F_{\frac{7}{2}}$ & 11.013 &11.017 &11.004\\
			$1^{2}G_{\frac{9}{2}}$ & 11.263 &11.267 &11.259\\
			$1^{2}H_{\frac{11}{2}}$& 11.508 &11.512\\
			\noalign{\smallskip}
			$1^{4}S_{\frac{3}{2}}$ & 10.330 &10.333 &10.316 &10.367 &10.352 &10.339\\
			$1^{4}P_{\frac{5}{2}}$ & 10.612 &10.615 &10.588 &10.731 &10.695 &10.759\\ 
			$1^{4}D_{\frac{7}{2}}$ & 10.886 &10.889 &10.853 &10.608 &11.011\\
			$1^{4}F_{\frac{9}{2}}$ & 11.154 &11.157 &11.112\\
			$1^{4}G_{\frac{11}{2}}$& 11.415 &11.418 &11.365\\
			$1^{4}H_{\frac{13}{2}}$& 11.670 &11.673 \\	
		\end{tabular}
	\end{ruledtabular}
\end{table}

\begin{table}%The best place to locate the table environment is directly after its first reference in text
	\caption{\label{tab:table10}
		Masses of excited states of $\Omega_{bb}^{-}$ baryon in the $(J,M^{2})$ plane for natural and unnatural parity states  (in GeV) .
	}
	\begin{ruledtabular}
		
		\begin{tabular}{lllllllllll}
			\textit{$N^{2S+1}L_{J}$} & Present & \cite{Zalak.bb2016} &\cite{Wei.2017} &\cite{Robert2008} &\cite{Ebert2005.2.8} &\cite{Yoshida2015}\\	
			\colrule\noalign{\smallskip}
			$1^{2}S_{\frac{1}{2}}$ & 10.350 &10.446 &10.320 &10.454 &10.359 &10.447\\
			$1^{2}P_{\frac{3}{2}}$ & 10.638 &10.641 &10.593 &10.619 &10.566 &10.608\\
			$1^{2}D_{\frac{5}{2}}$ & 10.918 &10.792 &10.858 &10.720 & &10.729\\
			$1^{2}F_{\frac{7}{2}}$ & 11.191 &10.930 &11.118\\
			$1^{2}G_{\frac{9}{2}}$ & 11.458 & &11.372\\
			$1^{2}H_{\frac{11}{2}}$& 11.718\\
			\noalign{\smallskip}
			$1^{4}S_{\frac{3}{2}}$ & 10.449 &10.467 &10.431 &10.486 &10.389 &10.467 \\
			$1^{4}P_{\frac{5}{2}}$ & 10.729 &10.637 &10.700 &10.766 &10.798 &10.808 \\ 
			$1^{4}D_{\frac{7}{2}}$ & 11.002 &10.786 &10.964 &10.732 \\
			$1^{4}F_{\frac{9}{2}}$ & 11.268 &10.924 &11.221 \\
			$1^{4}G_{\frac{11}{2}}$& 11.528 & &11.472 \\
			$1^{4}H_{\frac{13}{2}}$& 11.782 \\	
		\end{tabular}
	\end{ruledtabular}
\end{table}

\begin{table}%The best place to locate the table environment is directly after its first reference in text
	\caption{\label{tab:table11}
		Masses of excited states of $\Omega_{bbb}^{*-}$ baryon in the $(J,M^{2})$ plane for unnatural parity states  (in GeV).
	}
	\begin{ruledtabular}
		\begin{tabular}{llllllllll}
			\textit{$N^{2S+1}L_{J}$} & Present & \cite{Zalak.bbb2017}  & \cite{Wei.2017} &\cite{Robert2008} &\cite{Z. Wang 2010.11} & \cite{S. Meinel} \\	
			\colrule\noalign{\smallskip}
			$1^{4}S_{\frac{3}{2}}$ & 14.822 &14.496 & 14.788&14.834 &14.830&14.371\\
			$1^{4}P_{\frac{5}{2}}$ & 15.097 &14.931 &      \\
			$1^{4}D_{\frac{7}{2}}$ & 15.367 &15.286 &15.318&15.101 & &14.969\\
			$1^{4}F_{\frac{9}{2}}$ & 15.632 &15.631 &\\
			$1^{4}G_{\frac{11}{2}}$& 15.893 & & 15.831\\
			$1^{4}H_{\frac{13}{2}}$& 16.150\\
		\end{tabular}
	\end{ruledtabular}
\end{table}

\begin{table*}%The best place to locate the table environment is directly after its first reference in text 	
	\caption{\label{tab:table12}
		Masses of excited states of $\Lambda_{b}^{0}$ baryon in $(n,M^{2})$ plane  (in GeV). The masses from Ref. \cite{Thakkar2017}  are taken as input.}
	
	\begin{ruledtabular}
		
		\begin{tabular}{lllllllllllllll}
			&	\textit{$N^{2S+1}L_{J}$}  & Present &	\cite{Ebert2011} & \cite{Yoshida2015} & \cite{Yamaguchi2015} & \cite{S. Capstick1986} & \cite{Robert2008} &	\\
			
			\noalign{\smallskip}\hline\noalign{\smallskip}
			(S=1/2)& $1^{2}S_{1/2}$ &5.620 &5.620 &5.618 &5.612&5.585&5.612 \\
			& $2^{2}S_{1/2}$ &\textbf{6.026} \cite{Thakkar2017}&6.089 & &6.107\\
			& $  3^{2}S_{1/2}$ &6.406 &6.455 & &6.338\\
			& $4^{2}S_{1/2}$ &6.765 &6.756\\
			& $5^{2}S_{1/2}$ &7.106 &7.015\\
			& $6^{2}S_{1/2}$ &7.431 &7.256\\
			
			\noalign{\smallskip}\hline\noalign{\smallskip}    
			
			(S=1/2)& $1^{2}P_{3/2}$ &5.924 &5.942 &5.939 & &5.920&5.941 \\
			& $2^{2}P_{3/2}$ &\textbf{6.304} \cite{Thakkar2017} &6.333 &6.273\\
			& $3^{2}P_{3/2}$ &6.662 &6.651 &6.285\\
			& $4^{2}P_{3/2}$ &7.002 &6.922\\
			& $5^{2}P_{3/2}$ &7.327 &7.171\\

			\noalign{\smallskip}\hline\noalign{\smallskip}    
			
			(S=1/2) & $1^{2}D_{5/2}$ &6.213 &6.196 &6.212 & &6.165&6.183\\
			& $2^{2}D_{5/2}$ &\textbf{6.527} \cite{Thakkar2017}&6.531 &6.530\\
			& $3^{2}D_{5/2}$ &6.826 &6.814\\
			& $4^{2}D_{5/2}$ &7.113 &7.063\\
			& $5^{2}D_{5/2}$ &7.389 &\\

		\end{tabular}
	\end{ruledtabular} 
\end{table*}

\begin{table*}%The best place to locate the table environment is directly after its first reference in text  	
	\caption{\label{tab:table13}	Masses of excited states of $\Sigma_{b}^{\pm}$ baryon in $(n,M^{2})$ plane (in GeV). The masses from Ref. \cite{Thakkar2017} are taken as input.}
	
	\begin{ruledtabular}
		
		\begin{tabular}{lllllllllllllll}
			&	\textit{$N^{2S+1}L_{J}$}&\multicolumn{2}{c}{Present}& & Others	\\
			\colrule\noalign{\smallskip}

			& &$\Sigma_{b}^{+}$	  & $\Sigma_{b}^{-}$ &	\cite{Ebert2011} & \cite{Yoshida2015} & \cite{Yamaguchi2015} & \cite{Wei.2017}  	\\
			
			\noalign{\smallskip}\hline\noalign{\smallskip}
			(S=1/2)& $1^{2}S_{1/2}$ &5.811&5.816 &5.808 &5.823 &5.833 &5.813\\
			& $2^{2}S_{1/2}$ &\textbf{6.275} \cite{Thakkar2017} &\textbf{6.262} \cite{Thakkar2017} &6.213 &6.343 &6.294\\
			& $3^{2}S_{1/2}$ &6.707 &6.678 &6.575 &6.395\\
			& $4^{2}S_{1/2}$ &7.113 &7.070 &6.869 & &6.447\\
			& $5^{2}S_{1/2}$ &7.497 &7.441 &7.124\\
			& $6^{2}S_{1/2}$ &7.862 &7.795 &\\
			(S=3/2)& $1^{4}S_{3/2}$ &5.830 & 5.835 &5.834 &5.845 & &5.833\\
			& $2^{4}S_{3/2}$ &\textbf{6.291} \cite{Thakkar2017} &\textbf{6.277} \cite{Thakkar2017}&6.226 &6.356 &6.326\\
			& $3^{4}S_{3/2}$ &6.720 &6.690 &6.583 &6.393\\
			& $4^{4}S_{3/2}$ &7.124 &7.079 &6.876 & &6.447\\
			& $5^{4}S_{3/2}$ &7.506 &7.447 &7.129\\
			& $6^{4}S_{3/2}$ &7.869 &7.798&\\   
			
			\noalign{\smallskip}\hline\noalign{\smallskip}    
			
			(S=1/2)& $1^{2}P_{3/2}$ &6.105 &6.110 &6.096 &6.132 & &6.098 \\
			& $2^{2}P_{3/2}$ &\textbf{6.506} \cite{Thakkar2017}&\textbf{6.484} \cite{Thakkar2017}&6.430 &6.141\\
			& $3^{2}P_{3/2}$ &6.884&6.837 &6.742 &6.246\\
			& $4^{2}P_{3/2}$ &7.242 &7.174 &7.009\\
			& $5^{2}P_{3/2}$ &7.583 &7.495\\
			
			(S=3/2) & $1^{4}P_{5/2}$ &6.118&6.123 &6.084 &6.144 & &6.117 \\
			& $2^{4}P_{5/2}$ &\textbf{6.489} \cite{Thakkar2017}&\textbf{6.468} \cite{Thakkar2017} &6.421 &6.592\\
			& $3^{4}P_{5/2}$ &6.840 &6.795 &6.732 &6.834\\
			& $4^{4}P_{5/2}$ &7.174 &7.108 &6.999\\
			& $5^{4}P_{5/2}$ &7.493&7.407 &\\
			
			\noalign{\smallskip}\hline\noalign{\smallskip}    
			
			(S=1/2) & $1^{2}D_{5/2}$ &6.386&6.390 &6.284 &6.397 & &6.396\\
			& $2^{2}D_{5/2}$ &\textbf{6.778} \cite{Thakkar2017}&\textbf{6.746} \cite{Thakkar2017} &6.612 &6.402\\
			& $3^{2}D_{5/2}$ &7.148 &7.084\\
			& $4^{2}D_{5/2}$ &7.501 &7.407\\
			& $5^{2}D_{5/2}$ &7.837 &7.716\\
			
			(S=3/2) & $1^{4}D_{7/2}$ &6.393 &6.398&6.260 & & &6.388\\
			& $2^{4}D_{7/2}$ &\textbf{6.751} \cite{Thakkar2017}&\textbf{6.721} \cite{Thakkar2017} &6.590\\
			& $3^{4}D_{7/2}$ &7.091 &7.029\\
			& $4^{4}D_{7/2}$ &7.415 &7.324\\
			& $5^{4}D_{7/2}$ &7.726 &7.608\\

		\end{tabular}
	\end{ruledtabular}
\end{table*}

\begin{table*}%The best place to locate the table environment is directly after its first reference in text
	\caption{\label{tab:table14}
		Masses of excited states of $\Xi_{b}^{0}$ and $\Xi_{b}^{-}$ baryon in $(n,M^{2})$ plane (in GeV). The masses from Ref. \cite{Thakkar2017} are taken as input.}
	
	\begin{ruledtabular}
		
		\begin{tabular}{llllllllllll}
			
			&\textit{$N^{2S+1}L_{J}$}&\multicolumn{2}{c}{Present}& & Others	\\
			\colrule\noalign{\smallskip}
			& &$\Xi_{b}^{0}$ & $\Xi_{b}^{-}$ & \cite  {Ebert2011} & \cite{Yamaguchi2015} & \cite{Robert2008} &\cite{Wei.2017} &\cite{H. Garcilazo2007}\\
			\noalign{\smallskip}\hline\noalign{\smallskip}
			(S=1/2)& $1^{2}S_{1/2}$ &5.792&5.797 &5.803 &5.806 &5.806 &5.793&5.825 \\
			& $2^{2}S_{1/2}$	&\textbf{6.203} \cite{Thakkar2017}&\textbf{6.189} \cite{Thakkar2017} &6.266 &6.230\\
			& $3^{2}S_{1/2}$ &6.588 &6.558 &6.601 &6.547\\
			& $4^{2}S_{1/2}$ &6.952&6.907 &6.913\\
			& $5^{2}S_{1/2}$ &7.298 &7.239&7.165\\
			& $6^{2}S_{1/2}$ &7.629&7.556 &\\
			(S=3/2)& $1^{4}S_{3/2}$ &5.952 &5.955&5.963 & & 5.980&5.952 &5.967\\
			& $2^{4}S_{3/2}$ &\textbf{6.316} \cite{Thakkar2017}&\textbf{6.298} \cite{Thakkar2017} &\\
			& $3^{4}S_{3/2}$ &6.660&6.623 &\\
			& $4^{4}S_{3/2}$ &6.987&6.933 &\\
			& $5^{4}S_{3/2}$ &7.300&7.230 &\\
			& $6^{4}S_{3/2}$ &7.600&7.515 &\\   
			
			\noalign{\smallskip}\hline\noalign{\smallskip}    
			
			(S=1/2)&$1^{2}P_{3/2}$ &6.093&6.098 &6.130 & &6.093 &6.080&6.076 \\
			& $2^{2}P_{3/2}$	&\textbf{6.460} \cite{Thakkar2017}&\textbf{6.437} \cite{Thakkar2017} &6.502\\
			& $3^{2}P_{3/2}$ &6.807&6.759 &6.810\\
			& $4^{2}P_{3/2}$ &7.138&7.066 &7.073\\
			& $5^{2}P_{3/2}$ &7.453&7.361 &7.306\\
			
			(S=3/2) & $1^{4}P_{5/2}$ &6.240 &6.243 & & &6.201&6.232\\
			& $2^{4}P_{5/2}$ &\textbf{6.451} \cite{Thakkar2017}&\textbf{6.428} \cite{Thakkar2017}&\\
			& $3^{4}P_{5/2}$ &6.655 &6.608\\
			& $4^{4}P_{5/2}$ &6.853 &6.783\\
			& $5^{4}P_{5/2}$ &7.046 &6.953\\
			
			\noalign{\smallskip}\hline\noalign{\smallskip}    
			
			(S=1/2) & $1^{2}D_{5/2}$ &6.380&6.385 &6.373 & & 6.300&6.354\\
			& $2^{2}D_{5/2}$ &\textbf{6.687} \cite{Thakkar2017}&\textbf{6.696} \cite{Thakkar2017} &6.655\\
			& $3^{2}D_{5/2}$ &6.980&6.993 &\\
			& $4^{2}D_{5/2}$ &7.262&7.278 &\\
			& $5^{2}D_{5/2}$ &7.533 &7.552\\
			
			(S=3/2) & $1^{4}D_{7/2}$ &6.516 &6.518 & & &6.395&6.499 \\
			& $2^{4}D_{7/2}$ &\textbf{6.672} \cite{Thakkar2017}&\textbf{6.661} \cite{Thakkar2017} &\\
			& $3^{4}D_{7/2}$ &6.824 &6.801\\
			& $4^{4}D_{7/2}$ &6.973 &6.938\\
			& $5^{4}D_{7/2}$ &7.119 &7.073\\
			
		\end{tabular}
	\end{ruledtabular}
\end{table*}

\begin{table*}%The best place to locate the table environment is directly after its first reference in text 	
	\caption{\label{tab:tablecas}
		Masses of excited states of $\Xi_{b}^{'-}$ baryon in $(n,M^{2})$ plane (in GeV). The masses from Ref. \cite{Thakkar2017} are taken as input.}
	
	\begin{ruledtabular}
		
		\begin{tabular}{lllllllllllllll}
			&\textit{$N^{2S+1}L_{J}$}  & Present & \cite{Wei.2017} &\cite{Robert2008} &\cite{H. Garcilazo2007}	\\
			
			\noalign{\smallskip}\hline\noalign{\smallskip}
			(S=1/2)& $1^{2}S_{1/2}$ &5.935 &5.935 &5.970 &5.913\\
			& $2^{2}S_{1/2}$ &\textbf{6.329} \cite{Thakkar2017}\\
			& $  3^{2}S_{1/2}$ &6.700\\
			& $4^{2}S_{1/2}$ &7.051\\
			& $5^{2}S_{1/2}$ &7.386\\
			& $6^{2}S_{1/2}$ &7.706&\\
			
			\noalign{\smallskip}\hline\noalign{\smallskip}    
			
			(S=1/2)& $1^{2}P_{3/2}$ &6.229 &6.215 &6.190 &6.157\\
			& $2^{2}P_{3/2}$ &\textbf{6.605} \cite{Thakkar2017}\\
			& $3^{2}P_{3/2}$ &6.961\\
			& $4^{2}P_{3/2}$ &7.299\\
			& $5^{2}P_{3/2}$ &7.622\\

			\noalign{\smallskip}\hline\noalign{\smallskip}    
			
			(S=1/2) & $1^{2}D_{5/2}$ &6.510 &6.486&6.393\\
			& $2^{2}D_{5/2}$ &\textbf{6.751} \cite{Thakkar2017}\\
			& $3^{2}D_{5/2}$ &6.984\\
			& $4^{2}D_{5/2}$ &7.209\\
			& $5^{2}D_{5/2}$ &7.427&\\

		\end{tabular}
	\end{ruledtabular} 
\end{table*} 
\begin{table*}%The best place to locate the table environment is directly after its first reference in text

	\caption{\label{tab:table15}
		Masses of excited states of $\Omega_{b}^{-}$ baryon in $(n,M^{2})$ plane (in GeV). The masses from Ref. \cite{Thakkar2017} are taken as input.}
	\begin{ruledtabular}
		
		\begin{tabular}{lllllllllllllll}
			&\textit{$N^{2S+1}L_{J}$} & Present & \cite  {Ebert2011} & \cite{Yoshida2015} & \cite{Yamaguchi2015} & \cite{S.Agaev2017} & \cite{Wei.2017} \\
			\noalign{\smallskip}\hline\noalign{\smallskip}
			(S=1/2)& $1^{2}S_{1/2}$ &6.054 &6.064 &6.076 &6.081 &6.024 &6.048\\
			& $2^{2}S_{1/2}$ &\textbf{6.455} \cite{Thakkar2017}&6.450 &6.517 &6.472 &6.325\\
			& $3^{2}S_{1/2}$ &6.832 &6.804 &6.561 &6.593\\
			& $4^{2}S_{1/2}$ &7.190 &7.091 & &6.648\\
			& $5^{2}S_{1/2}$ &7.531 &7.338\\
			& $6^{2}S_{1/2}$ &7.857 &\\
			(S=3/2)& $1^{4}S_{3/2}$ &6.074 &6.088 & &6.102 &6.084 &6.069\\
			& $2^{4}S_{3/2}$ &\textbf{6.481} \cite{Thakkar2017} &6.461 &6.528 &6.478 &6.412\\
			& $3^{4}S_{3/2}$ &6.864 &6.811 &6.559 &6.593\\
			& $4^{4}S_{3/2}$ &7.226 &7.096 & &6.645\\
			& $5^{4}S_{3/2}$ &7.572 &7.343\\
			& $6^{4}S_{3/2}$ &7.902 &\\   
			
			\noalign{\smallskip}\hline\noalign{\smallskip}    
			
			(S=1/2)& $1^{2}P_{3/2}$ &6.348 &6.340 &6.336 & & & 6.325\\
			& $2^{2}P_{3/2}$ &\textbf{6.662} \cite{Thakkar2017} &6.705 &6.344\\
			& $3^{2}P_{3/2}$ &6.962 &7.002 &6.919\\
			& $4^{2}P_{3/2}$ &7.249 &7.258\\
			& $5^{2}P_{3/2}$ &7.526\\
			
			(S=3/2) & $1^{4}P_{5/2}$ &6.362 &6.334 & & & &6.345\\
			& $2^{4}P_{5/2}$ &\textbf{6.653} \cite{Thakkar2017}&6.700\\
			& $3^{4}P_{5/2}$ &6.932 &6.996\\
			& $4^{4}P_{5/2}$ &7.200 &7.251\\
			& $5^{4}P_{5/2}$ &7.458 &\\
			
			\noalign{\smallskip}\hline\noalign{\smallskip}    
			
			(S=1/2) & $1^{2}D_{5/2}$ &6.629  &6.529 &6.561 & & &6.590\\
			& $2^{2}D_{5/2}$ &\textbf{6.659} \cite{Thakkar2017}&6.846 &6.566\\
			& $3^{2}D_{5/2}$ &6.689  &\\
			& $4^{2}D_{5/2}$ &6.719 &\\
			& $5^{2}D_{5/2}$ &6.748 &\\
			
			(S=3/2) & $1^{4}D_{7/2}$ &6.638 &6.517 & & & & 6.609\\
			& $2^{4}D_{7/2}$ &\textbf{6.643} \cite{Thakkar2017}&6.834\\
			& $3^{4}D_{7/2}$ &6.648 &\\
			& $4^{4}D_{7/2}$ &6.653 &\\
			& $5^{4}D_{7/2}$ &6.658 &\\
			
		\end{tabular}
	\end{ruledtabular}
\end{table*}

\begin{table*}%The best place to locate the table environment is directly after its first reference in text
	\caption{\label{tab:table16}
		Masses of excited states of $\Xi_{bb}^{0}$ and $\Xi_{bb}^{-}$ baryon in $(n,M^{2})$ plane (in GeV).  The masses from Ref. \cite{Zalak.bb2016} are taken as input.
	}
	\begin{ruledtabular}
		
		\begin{tabular}{lllllllllllllll}
			&\textit{$N^{2S+1}L_{J}$}&\multicolumn{2}{c}{Present}& & Others	\\
			\colrule\noalign{\smallskip}
			&  & $\Xi_{bb}^{0}$ &$\Xi_{bb}^{-}$&\cite{Yoshida2015} & \cite{Robert2008} &\cite{Ebert2005.2.8}  &\cite{F.Giannuzzi2009} & \cite{B. Eakins2012} \\
			\noalign{\smallskip}\hline\noalign{\smallskip}
			(S=1/2)& $1^{2}S_{1/2}$ &10.225 &10.230&10.314 &10.340 &10.202 &10.185 &10.322\\
			& $2^{2}S_{1/2}$ &\textbf{10.609} \cite{Zalak.bb2016} &\textbf{10.612} \cite{Zalak.bb2016}&10.571 &10.576 &10.441 &10.751 &10.551\\
			& $3^{2}S_{1/2}$ &10.979 &10.981&10.612 & &10.630 &11.170\\
			& $4^{2}S_{1/2}$ &11.338 &11.337& & &10.812\\
			& $5^{2}S_{1/2}$ &11.685&11.683\\
			& $6^{2}S_{1/2}$ &12.023&12.019 &\\
			(S=3/2)& $1^{4}S_{3/2}$ &10.330&10.333 &10.339 &10.367 &10.237 &10.216 &10.352\\
			& $2^{4}S_{3/2}$ &\textbf{10.617} \cite{Zalak.bb2016}&\textbf{10.619} \cite{Zalak.bb2016} &10.592 &10.578 &10.428 &10.770 &10.574\\
			& $3^{4}S_{3/2}$ &10.896&10.897 &10.593 & &10.673 &11.184\\
			& $4^{4}S_{3/2}$ &11.169&11.169 & & &10.856\\
			& $5^{4}S_{3/2}$ &11.435&11.434\\
			& $6^{4}S_{3/2}$ &11.695 &11.693\\   
			
			\noalign{\smallskip}\hline\noalign{\smallskip}    
			
			(S=1/2)& $1^{2}P_{3/2}$ &10.494&10.499 &10.476 &10.495 &10.408 & &10.692\\
			& $2^{2}P_{3/2}$ &\textbf{10.765} \cite{Zalak.bb2016}&\textbf{10.766} \cite{Zalak.bb2016} &10.704 &10.713 &10.607\\
			& $3^{2}P_{3/2}$ &11.029&11.026 &10.742 & &10.788\\
			& $4^{2}P_{3/2}$ &11.287&11.281\\
			& $5^{2}P_{3/2}$ &11.540&11.530\\
			
			(S=3/2) & $1^{4}P_{5/2}$ &10.612&10.615 &10.759 & & & &10.695\\
			& $2^{4}P_{5/2}$ &\textbf{10.776} \cite{Zalak.bb2016}&\textbf{10.763} \cite{Zalak.bb2016}&10.973 &10.713\\
			& $3^{4}P_{5/2}$ &10.937&10.909 &11.004\\
			& $4^{4}P_{5/2}$ &11.097&11.053\\
			& $5^{4}P_{5/2}$ &11.254&11.195\\
			
			\noalign{\smallskip}\hline\noalign{\smallskip}    
			
			(S=1/2) & $1^{2}D_{5/2}$ &10.757&10.761 &10.592 &10.676 & & &11.002\\
			& $2^{2}D_{5/2}$ &\textbf{10.901} \cite{Zalak.bb2016}&\textbf{10.901} \cite{Zalak.bb2016}& &10.712\\
			& $3^{2}D_{5/2}$ &11.043&11.039 &\\
			& $4^{2}D_{5/2}$ &11.183&11.176\\
			& $5^{2}D_{5/2}$ &11.322 &11.311\\
			
			(S=3/2) & $1^{4}D_{7/2}$ &10.886&10.889 & &10.608 & & &11.011 \\
			& $2^{4}D_{7/2}$ &\textbf{10.896} \cite{Zalak.bb2016}&\textbf{10.896} \cite{Zalak.bb2016}& &11.057\\
			& $3^{4}D_{7/2}$ &10.906&10.903 &\\
			& $4^{4}D_{7/2}$ &10.916&10.910 &\\
			& $5^{4}D_{7/2}$ &10.926&10.917 &\\
			
		\end{tabular}
	\end{ruledtabular}
\end{table*}

\begin{table*}%The best place to locate the table environment is directly after its first reference in text
	\caption{\label{tab:table17}
		Masses of excited states of $\Omega_{bb}^{-}$ baryon in $(n,M^{2})$ plane (in GeV). The masses from Ref. \cite{Zalak.bb2016} are taken as input.
	}
	\begin{ruledtabular}
		
		\begin{tabular}{lllllllllllllll}
			&\textit{$N^{2S+1}L_{J}$} & Present &\cite{Yoshida2015} & \cite{Robert2008} &\cite{Ebert2005.2.8}  &\cite{F.Giannuzzi2009} & \cite{Valcarce2008}\\
			\noalign{\smallskip}\hline\noalign{\smallskip}
			(S=1/2)& $1^{2}S_{1/2}$ &10.350 &10.447 &10.454 &10.359 &10.271 &10.293\\
			& $2^{2}S_{1/2}$ &\textbf{10.736} \cite{Zalak.bb2016} &10.707 &10.693 &10.610 &10.830 &10.604\\
			& $3^{2}S_{1/2}$ &11.109 &10.744 & &10.806 &11.240\\
			& $4^{2}S_{1/2}$ &11.469 &10.994 & &\\
			& $5^{2}S_{1/2}$ &11.819 & & &\\
			& $6^{2}S_{1/2}$ &12.158 &\\
			(S=3/2)& $1^{4}S_{3/2}$ &10.449 &10.467 &10.486 &10.389 &10.289 &10.321\\
			& $2^{4}S_{3/2}$ &\textbf{10.743} \cite{Zalak.bb2016}&10.723 &10.721 &10.645 &10.839 &10.622\\
			& $3^{4}S_{3/2}$ &11.029 &10.730 & &10.843 &11.247\\
			& $4^{4}S_{3/2}$ &11.308 &11.031 & &\\
			& $5^{4}S_{3/2}$ &11.580 & & &\\
			& $6^{4}S_{3/2}$ &11.846 & & &\\   
			
			\noalign{\smallskip}\hline\noalign{\smallskip}    
			
			(S=1/2)& $1^{2}P_{3/2}$ &10.638 &10.608 &10.619 &10.566\\
			& $2^{2}P_{3/2}$ &\textbf{10.893} \cite{Zalak.bb2016}&10.797 &10.765 &10.775\\
			& $3^{2}P_{3/2}$ &11.142 &10.805\\
			& $4^{2}P_{3/2}$ &11.386\\
			& $5^{2}P_{3/2}$ &11.624\\
			
			(S=3/2) & $1^{4}P_{5/2}$ &10.729 &10.808 &10.766 &10.798\\
			& $2^{4}P_{5/2}$ &\textbf{10.888} \cite{Zalak.bb2016}&11.028\\
			& $3^{4}P_{5/2}$ &11.045 &11.059\\
			& $4^{4}P_{5/2}$ &11.199\\
			& $5^{4}P_{5/2}$ &11.352\\
			
			\noalign{\smallskip}\hline\noalign{\smallskip}    
			
			(S=1/2) & $1^{2}D_{5/2}$ &10.918 &10.729 &10.720\\
			& $2^{2}D_{5/2}$ &\textbf{11.025} \cite{Zalak.bb2016} &10.744 &10.734\\
			& $3^{2}D_{5/2}$ &11.131&10.937\\
			& $4^{2}D_{5/2}$ &11.236&\\
			& $5^{2}D_{5/2}$ &11.340 &\\
			
			(S=3/2) & $1^{4}D_{7/2}$ &11.002 \\
			& $2^{4}D_{7/2}$ &\textbf{11.021} \cite{Zalak.bb2016}\\
			& $3^{4}D_{7/2}$ &11.040 &\\
			& $4^{4}D_{7/2}$ &11.059 &\\
			& $5^{4}D_{7/2}$ &11.078 &\\
			
		\end{tabular}
	\end{ruledtabular}
\end{table*}

\begin{table}%The best place to locate the table environment is directly after its first reference in text
	\caption{\label{tab:table18}
		Masses of excited states of $\Omega_{bbb}^{*-}$ baryon in $(n,M^{2})$ plane (in GeV). The masses from Ref. \cite{Zalak.bbb2017} are taken as input.
	}
	\begin{ruledtabular}
		
		\begin{tabular}{lllllllllllllll}
			& \textit{$N^{2S+1}L_{J}$} & Present &\cite{Wei.2017} & \cite{S. Meinel} &\cite{Robert2008}\\
			\noalign{\smallskip}\hline\noalign{\smallskip}
			
			(S=3/2)& $1^{4}S_{3/2}$ &14.822 &14.788 & 14.371 &14.834\\
			& $2^{4}S_{3/2}$ &\textbf{15.163} \cite{Zalak.bbb2017}\\
			& $3^{4}S_{3/2}$ &15.496\\
			& $4^{4}S_{3/2}$ &15.823&\\
			& $5^{4}S_{3/2}$ &16.143\\
			& $6^{4}S_{3/2}$ &16.456\\   
			
			\noalign{\smallskip}\hline\noalign{\smallskip}

			(S=3/2) & $1^{4}P_{5/2}$ &15.097\\
			& $2^{4}P_{5/2}$ &\textbf{15.425} \cite{Zalak.bbb2017}\\
			& $3^{4}P_{5/2}$ &15.746\\
			& $4^{4}P_{5/2}$ &16.061\\
			& $5^{4}P_{5/2}$ &16.369\\
			
			\noalign{\smallskip}\hline\noalign{\smallskip}

			(S=3/2) & $1^{4}D_{7/2}$ &15.367 &15.318 &14.969 &15.101\\
			& $2^{4}D_{7/2}$ &\textbf{15.776} \cite{Zalak.bbb2017}\\
			& $3^{4}D_{7/2}$ &16.175 &\\
			& $4^{4}D_{7/2}$ &16.564 &\\
			& $5^{4}D_{7/2}$ &16.944 &\\
			
		\end{tabular}
	\end{ruledtabular}
\end{table}

\subsection{Ground state masses of $\Omega_{b}^{-}$, $\Xi_{bb}^{0,-}$, $\Omega_{bb}^{-}$ and $\Omega_{bbb}^{*-}$ baryons}

Eq. (\ref{eq:5}) we obtain above can also be expressed as,
\begin{equation}
	\label{eq:17}
	\dfrac{\alpha^{'}_{jjq}}{\alpha^{'}_{iiq}}=	\dfrac{\alpha^{'}_{kkq}}{\alpha^{'}_{iiq}}\times	\dfrac{\alpha^{'}_{jjq}}{\alpha^{'}_{kkq}} ,
\end{equation}	
here $k$ can be any quark flavor. Thus we have,
\begin{widetext}
	\begin{equation}
		\label{eq:18}
		\begin{split}
			\dfrac{[(4M^{2}_{ijq}-M^{2}_{iiq}-M^{2}_{jjq})+\sqrt{{{(4M^{2}_{ijq}-M^{2}_{iiq}-M^{2}_{jjq}})^2}-4M^{2}_{iiq}M^{2}_{jjq}}]}{2M^{2}_{jjq}} \\
			=\dfrac{[(4M^{2}_{ikq}-M^{2}_{iiq}-M^{2}_{kkq})+\sqrt{{{(4M^{2}_{ikq}-M^{2}_{iiq}-M^{2}_{kkq}})^2}-4M^{2}_{iiq}M^{2}_{kkq}}]/2M^{2}_{kkq}}{[(4M^{2}_{jkq}-M^{2}_{jjq}-M^{2}_{kkq})+\sqrt{{{(4M^{2}_{jkq}-M^{2}_{jjq}-M^{2}_{kkq}})^2}-4M^{2}_{jjq}M^{2}_{kkq}}]/2M^{2}_{kkq}} .
		\end{split}
	\end{equation}
\end{widetext}
This is the general relation we have derived in terms of baryon masses which can be used to predict the mass of any baryon state if all other masses are known.
In the present work with the help of relation (\ref{eq:18}), we evaluate the ground  state masses of unobserved bottom baryons. Since $\Omega_{b}^{-}$ comprises two $s$-quarks and one $b$-quark. When we put $i=n$ ($u$ or $d$), $j=s$, $q=b$, and $k=n$ in Eq. (\ref{eq:18}) we have,

\begin{eqnarray}
	\label{eq:19}
	\begin{split}
	& \left[(M_{\Sigma_{b}^{-}}+M_{\Omega_{b}^{-}})^{2}-4M^{2}_{\Xi_{b}^{'-}}\right] \\
	& =
	 \sqrt{(4M^{2}_{\Xi_{b}^{'-}}-M^{2}_{\Sigma_{b}^{-}}-M^{2}_{\Omega_{b}^{
	 			-}})^{2}-4M^{2}_{\Sigma_{b}^{-}}M^{2}_{\Omega_{b}^{-}}}
	\end{split}
\end{eqnarray}
inserting the masses of $\Sigma_{b}^{-}$ and $\Xi_{b}^{'-}$ ($J^{P}=\frac{1}{2}^{+}$) from PDG \cite{PDG} into Eq. (\ref{eq:19}), we obtain the ground state mass of $\Omega_{b}^{-}$ baryon as 6.054 GeV for $J^{P}=\frac{1}{2}^{+}$. Similarly we get $M_{\Omega_{b}^{*-}}$ = 6.074 GeV for  $J^{P}=\frac{3}{2}^{+}$ state. In the same manner for doubly bottom baryons, in the case of $\Xi_{bb}^{0}$ which is composed of ($ubb$) we put $i=d$, $j=b$, $q=u$, $k=s$ and for $\Omega_{bb}^{-}$ which is composed of ($sbb$) we put $i=u$, $j=s$, $q=s$, $k=b$ in Eq. (\ref{eq:18}) and obtained the mass expressions to calculate the ground state masses of doubly bottom baryons $\Xi_{bb}^{0}$ and $\Omega_{bb}^{*-}$, which are expressed as a function of well-established masses of light baryons and singly bottom baryons.

\begin{widetext}
	For $\Xi_{bb}^{0}$;
	
	\begin{eqnarray}
		\begin{split}
		\label{eq:20}
	\dfrac{\left[(4M^{2}_{\Lambda_{b}^{0}}-M^{2}_{n}-M^{2}_{\Xi_{bb}^{0}})+\sqrt{(4M^{2}_{\Lambda_{b}^{0}}-M^{2}_{n}-M^{2}_{\Xi_{bb}^{0}})^{2}-4M^{2}_{n}M^{2}_{\Xi_{bb}^{0}}}\right]}{2M^{2}_{\Xi_{bb}^{0}}}\\ 
	=\frac{\left[(4M^{2}_{\Sigma^{0}}-M^{2}_{n}-M^{2}_{\Xi^{0}})+\sqrt{(4M^{2}_{\Sigma^{0}}-M^{2}_{n}-M^{2}_{\Xi^{0}})^{2}-4M^{2}_{n}M^{2}_{\Xi^{0}}}\right]}{\left[(4M^{2}_{\Xi_{b}^{0}}-M^{2}_{\Xi_{bb}^{0}}-M^{2}_{\Xi^{0}})+\sqrt{(4M^{2}_{\Xi_{b}^{0}}-M^{2}_{\Xi_{bb}^{0}}-M^{2}_{\Xi^{0}})^{2}-4M^{2}_{\Xi_{bb}^{0}}M^{2}_{\Xi^{0}}}\right]}
\end{split}
	\end{eqnarray}
	For $\Omega_{bb}^{*-}$; 
	
\begin{eqnarray}
	\begin{split}
	\label{eq:21}
	\dfrac{\left[(4M^{2}_{\Xi^{*0}}-M^{2}_{\Sigma^{*+}}-M^{2}_{\Omega^{-}})+\sqrt{(4M^{2}_{\Xi^{*0}}-M^{2}_{\Sigma^{*+}}-M^{2}_{\Omega^{-}})^{2}-4M^{2}_{\Sigma^{*+}}M^{2}_{\Omega^{-}}}\right]}{2M^{2}_{\Omega^{-}}}\\
	=\frac{\left[(4M^{2}_{\Xi_{b}^{*0}}-M^{2}_{\Sigma^{*+}}-M^{2}_{\Omega_{bb}^{*-}})+\sqrt{(4M^{2}_{\Xi_{b}^{*0}}-M^{2}_{\Sigma^{*+}}-M^{2}_{\Omega_{bb}^{*-}})^{2}-4M^{2}_{\Sigma^{*+}}M^{2}_{\Omega_{bb}^{*-}}}\right]}{\left[(4M^{2}_{\Omega_{b}^{*-}}-M^{2}_{\Omega^{-}}-M^{2}_{\Omega_{bb}^{*-}})+\sqrt{(4M^{2}_{\Omega_{b}^{*-}}-M^{2}_{\Omega^{-}}-M^{2}_{\Omega_{bb}^{*-}})^{2}-4M^{2}_{\Omega_{bb}^{*-}}M^{2}_{\Omega^{-}}}\right]}
\end{split}
	\end{eqnarray}

\end{widetext}

Inserting the masses of $\Lambda_{b}^{0}$, neutron $(n)$, $\Sigma^{0}$, $\Xi^{0}$, and $\Xi_{b}^{0}$ baryons from PDG \cite{PDG} in Eq. (\ref{eq:20}), we get $M_{\Xi_{bb}^{0}}$ = 10.225 GeV and also we can calculate $M_{\Xi_{bb}^{-}}$ = 10.230 GeV, for $J^{P}=\frac{1}{2}^{+}$ state. Similarly putting the masses of $\Xi^{*0}$, $\Sigma^{*+}$, $\Omega^{-}$, $\Xi_{b}^{*0}$ from \cite{PDG} and $\Omega_{b}^{*-}$ (calculated above) into Eq. (\ref{eq:21}), we get $M_{\Omega_{bb}^{*-}}$ = 10.449 GeV for $J^{P}=\frac{3}{2}^{+}$ state. Here in the calculation, we avoid using the masses of $\Delta^{++}$, $\Delta^{+}$, $\Delta^{0}$, and $\Delta^{-}$ baryons because only the charge mixed states of $\Delta(1232)$ were assuredly measured, as mentioned in PDG \cite{PDG}. Now since it is already stated in the above relation (\ref{eq:16}) that $\delta^{b}_{ij,q}$ is independent of quark flavor $q$. Therefore with the help of Eq. (\ref{eq:19}), we have the following relations:\\
(I) $i=u$, $j=s$, $q=u$, $b$
\begin{eqnarray}
	\label{eq:22}
	               \nonumber	
		\delta_{us}^{(3/2)^{+}}&=&M^{2}_{\Delta}+M^{2}_{\Xi^{*}}-2M^{2}_{\Sigma^{*}}\\
	&=&M^{2}_{\Sigma_{b}^{*}}+M^{2}_{\Omega_{b}^{*}}-2M^{2}_{\Xi_{b}^{*}};
\end{eqnarray}
(II) $i=u$, $j=b$, $q=u$ ,$s$
\begin{eqnarray}
	\label{eq:23}
	\nonumber
   \delta_{ub}^{(3/2)^{+}}&=&M^{2}_{\Delta}+M^{2}_{\Xi^{*}_{bb}}-2M^{2}_{\Sigma^{*}_{b}}\\
		&=&M^{2}_{\Sigma^{*}}+M^{2}_{\Omega_{bb}^{*}}-2M^{2}_{\Xi_{b}^{*}};
	\end{eqnarray}
(III) $i=s$, $j=b$, $q=u$ ,$b$
\begin{eqnarray}
	\label{eq:24}
	\nonumber
		\delta_{sb}^{(3/2)^{+}}&=&M^{2}_{\Xi^{*}}+M^{2}_{\Xi^{*}_{bb}}-2M^{2}_{\Xi^{*}_{b}}\\
		&=&M^{2}_{\Omega_{b}^{*}}+M^{2}_{\Omega_{bbb}^{*}}-2M^{2}_{\Omega_{bb}^{*}};
	\end{eqnarray}
solving Eqs. (\ref{eq:22}) and (\ref{eq:23}) we have,

\begin{equation}
\label{eq:25}
(M^{2}_{\Omega_{b b}^{* -}}-M^{2}_{\Xi^{* 0}_{b b}})
+ (M^{2}_{\Xi^{* 0}}-M^{2}_{\Sigma^{* +}})
= (M^{2}_{\Omega_{b}^{* -}}-M^{2}_{\Sigma_{b}^{* +}})
\end{equation}

similarly, its corresponding relation for $\frac{1}{2}^{+}$ multiplet is expressed as,
\begin{equation}
	\label{eq:26}
	(M^{2}_{\Omega_{bb}^{-}}-M^{2}_{\Xi_{bb}^{0}})	+ (M^{2}_{\Xi^{0}}-M^{2}_{\Sigma^{+}}) = (M^{2}_{\Omega_{b}^{-}}-M^{2}_{\Sigma_{b}^{+}})
\end{equation}

Now using relations (\ref{eq:25}) and (\ref{eq:26}), we can obtain the mass expressions for $\Xi_{bb}^{*0}$, $\Xi_{bb}^{*-}$  and $\Omega_{bb}^{-}$ baryons. Again after substituting all other masses, we get the ground-state masses $M_{\Xi_{bb}^{*0}}$ = 10.330 GeV,$M_{\Xi_{bb}^{*-}}$ = 10.333 GeV,  and $M_{\Omega_{bb}^{-}}$ = 10.350 GeV. Further to calculate the ground-state mass of $\Omega_{bbb}^{*-}$ baryon, we extract the mass expression with the help of Eqs. (\ref{eq:21}) and (\ref{eq:24}), which is expressed as,

\begin{widetext}
	{\tiny
		\begin{equation}
			\begin{split}
				\dfrac{\left(6M^{2}_{\Xi_{b}^{0}}-2M^{2}_{\Sigma^{+}}-M^{2}_{\Omega_{b}^{-}}-M^{2}_{\Omega_{bbb}^{-}}+M^{2}_{\Xi^{0}}+M^{2}_{\Xi_{bb}^{0}}\right) + \sqrt{\left(6M^{2}_{\Xi_{b}^{0}}-2M^{2}_{\Sigma^{+}}-M^{2}_{\Omega_{b}^{-}}-M^{2}_{\Omega_{bbb}^{-}}+M^{2}_{\Xi^{0}}+M^{2}_{\Xi_{bb}^{0}}\right)^{2}-8M^{2}_{\Sigma^{+}}\left(M^{2}_{\Omega_{b}^{-}}+M^{2}_{\Omega_{bbb}^{-}}-M^{2}_{\Xi^{0}}-M^{2}_{\Xi_{bb}^{0}}+2M^{2}_{\Xi_{b}^{0}}\right)}}{\left(4M^{2}_{\Xi^{0}}-M^{2}_{\Sigma^{+}}-M^{2}_{\Omega^{-}}\right)+\sqrt{\left(4M^{2}_{\Xi^{0}}-M^{2}_{\Sigma^{+}}-M^{2}_{\Omega^{-}}\right)^{2}-4M^{2}_{\Sigma^{+}}M^{2}_{\Omega^{-}}}}\\
				= \dfrac{\left(7M^{2}_{\Omega_{b}^{-}}-2M^{2}_{\Omega^{-}}-M^{2}_{\Omega_{bbb}^{-}}+M^{2}_{\Xi^{0}}+M^{2}_{\Xi_{bb}^{0}}-2M^{2}_{\Xi_{b}^{0}}\right)+\sqrt{\left(7M^{2}_{\Omega_{b}^{-}}-2M^{2}_{\Omega^{-}}-M^{2}_{\Omega_{bbb}^{-}}+M^{2}_{\Xi^{0}}+M^{2}_{\Xi_{bb}^{0}}-2M^{2}_{\Xi_{b}^{0}}\right)^{2}-8M^{2}_{\Omega^{-}}\left(M^{2}_{\Omega_{b}^{-}}+M^{2}_{\Omega_{bbb}^{-}}-M^{2}_{\Xi^{0}}-M^{2}_{\Xi_{bb}^{0}}+2M^{2}_{\Xi_{b}^{0}}\right)}}{2M^{2}_{\Omega^{-}}} ,
	\end{split}	
		\end{equation}
	}
\end{widetext}
 by substituting the masses of $\Xi_{b}^{0}$, $\Sigma^{+}$, $\Omega_{b}^{-}$, $\Xi^{0}$, $\Omega^{-}$, and $\Xi_{bb}^{0}$ ($J^{P}=\frac{3}{2}^{+}$),  we get $M_{\Omega_{bbb}^{*-}}$ = 14.822 GeV.
Table \ref{tab:table2} shows the comparison for the masses of $\Omega_{b}^{-}$, $\Xi_{bb}^{0}$,   $\Xi_{bb}^{-}$, $\Omega_{bb}^{-}$, and $\Omega_{bbb}^{*-}$ baryons with $J^{P}$ = $\frac{1}{2}^{+}$ and $\frac{3}{2}^{+}$ estimated in the present work and those predicted in other references.

\subsection{Excited state masses of singly, doubly and triply bottom baryons}

After evaluating the ground-state masses of unseen singly, doubly and triply bottom baryons, in this section we calculate the excited state masses of bottom baryons lying on $\frac{1}{2}^{+}$ and $\frac{3}{2}^{+}$ trajectories by obtaining the Regge slopes $\alpha^{'}$. For instance, using Eq. (\ref{eq:6}) we have,   
\begin{equation}
	\begin{split}
		\label{eq:27}
	\dfrac{\alpha^{'*}_{nsb}}{\alpha^{'}_{nns}}=\dfrac{1}{4M^{2}_{\Xi_{b}^{*}}}\times[(4M^{2}_{\Xi_{b}^{*}}+M^{2}_{\Sigma^{*}}-M^{2}_{\Omega_{bb}^{*}})\\
	+\sqrt{{{(4M^{2}_{\Xi_{b}^{*}}-M^{2}_{\Sigma^{*}}-M^{2}_{\Omega_{bb}^{*}}})^2}-4M^{2}_{\Sigma^{*}}M^{2}_{\Omega_{bb}^{*}}}],
\end{split}	
\end{equation}
Putting the values of masses in the above equation, we get $\alpha^{'}_{nsb}/\alpha^{'}_{nns}$. With the help of Eq. (\ref{eq:1}), we have
\begin{equation}
	\label{eq:28}
	\alpha^{'} = \dfrac{(J+2)-J}{M^{2}_{J+2}-M^{2}_{J}} ,
\end{equation} 
from above relation we have, $\alpha^{'*}_{nns}$ = 0.9057 GeV$^{-2}$. So we get $\alpha^{'*}_{nsb}$ = 0.2846 GeV$^{-2}$ for $\frac{3}{2}^{+}$ trajectory. Similarly with the aid of Eqs. (\ref{eq:3}), (\ref{eq:5}), and (\ref{eq:6}) we can find $\alpha^{'}_{nnb}$, $\alpha^{'}_{ssb}$, $\alpha^{'}_{nbb}$, $\alpha^{'}_{sbb}$, and $\alpha^{'}_{bbb}$ for both $\frac{1}{2}^{+}$ and $\frac{3}{2}^{+}$ trajectories. According to the Ref. \cite{L. Burakovsky}, $\alpha^{'}_{\Lambda_{b}} \simeq \alpha^{'}_{\Sigma_{b}}$ and $\alpha^{'}_{\Xi_{b}^{'}} \simeq \alpha^{'}_{\Xi_{b}}$. In this work, we take this approximation. The extracted Regge slopes $\alpha^{'}$ and $\alpha^{*'}$ for the bottom baryons in the present work for $\frac{1}{2}^{+}$ and $\frac{3}{2}^{+}$ trajectories respectively are shown in Table \ref{tab:table3}. In Ref. \cite{Wei.2017}, the authors gave the values of the Regge slopes for singly, doubly, and triply bottom baryons, which are approximately same to the corresponding values in this work. For example, $\alpha^{'}_{nnb}$ = 0.295$\pm$0.022 GeV$^{-2}$ in Ref. \cite{Wei.2017}, while $\alpha^{'}_{nnb}$ = 0.2852 GeV$^{-2}$ in the present work.

Now from Eq. (\ref{eq:1}) one can have,

\begin{equation}
	\label{eq:29}
	M_{J+1} = \sqrt{M_{J}^{2}+\dfrac{1}{\alpha^{'}}} .
\end{equation}
using the Regge slope $\alpha^{'}$, we obtain the orbitally excited state masses of singly, doubly and triply bottom baryons for both natural ($J^{P}$ = $1/2^{+}$, $3/2^{-}$, $5/2^{+}$, ....) and unnatural ($J^{P}$ = $3/2^{+}$, $5/2^{-}$, $7/2^{+}$, ....)  parities in the ($J,M^{2}$) plane with the help of Eq.(\ref{eq:29}). Table \ref{tab:table4} to \ref{tab:table11} shows the estimated results for singly, doubly, and triply bottom baryons in the present work and those of other theoretical approaches. Here the spectroscopic notations \textit{$N^{2S+1}L_{J}$}, is used to represent the state of the particles, where $N$, $L$, $S$ denotes the radial excited quantum number, orbital quantum number, and intrinsic spin, respectively

\subsection{Masses of singly, doubly, and triply bottom baryons in the ($n,M^{2}$) plane}
In this section, we evaluate the Regge parameters in the ($n,M^{2}$) plane to calculate the orbital and radial excited states of singly, doubly, and triply bottom baryons. The general equation for linear Regge trajectories in the ($n,M^{2}$) plane can be expressed as,
\begin{equation}
	\label{eq:30}
	n = \beta_{0} + \beta M^{2},
\end{equation}
where $n$ = 1, 2, 3.... is the radial principal quantum number, $\beta_{0}$, and $\beta$ are the intercept and slope of the trajectories. The baryon multiplets lying on the single Regge line have the same Regge slope ($\beta$) and Regge intercept ($\beta_{0}$). Using relation (\ref{eq:30}), we calculate $\beta$ and $\beta_{0}$ and with the help of these parameters we estimated the excited state masses of singly, doubly, and triply bottom baryons lying on each Regge lines for natural and unnatural parity states. For instance, using the slope equation, we have $\beta_{(S)} = 1/(M^{2}_{\Omega_{b}(2S)}-M^{2}_{\Omega_{b}(1S)})$ for $\Omega_{b}^{-}$ baryon, where $M_{\Omega_{b}(1S)}$ = 6.054 GeV (calculated above) and taking $M_{\Omega_{b}(2S)}$ = 6.455 GeV from \cite{Thakkar2017} for the $1/2^{+}$ trajectory, we get $\beta_{(S)}$ = 0.19935 GeV$^{-2}$. From Eq. (\ref{eq:30}) we can write,
\begin{equation}
	\label{eq:31}
	\begin{split}
		1 = \beta_{0(S)} + \beta_{(S)} M^{2}_{\Omega_{b}(1S)},\\
		2 = \beta_{0(S)} + \beta_{(S)} M^{2}_{\Omega_{b}(2S)},
	\end{split}
\end{equation}
using the above relations, we get $\beta_{0(S)}$ = -6.30664. With the help of $\beta_{(S)}$ and $\beta_{0(S)}$, we calculate the masses of the excited $\Omega_{b}^{-}$ baryon for $n$ = 3, 4, 5... Similarly, we can express these relations for $P$ and $D$-wave as,
\begin{equation}
	\label{eq:32}
	\begin{split}
		1 = \beta_{0(P)} + \beta_{(P)} M^{2}_{\Omega_{b}(1P)},\\
		2 = \beta_{0(P)} + \beta_{(P)} M^{2}_{\Omega_{b}(2P)},\\
		1 = \beta_{0(D)} + \beta_{(D)} M^{2}_{\Omega_{b}(1D)},\\
		2 = \beta_{0(D)} + \beta_{(D)} M^{2}_{\Omega_{b}(2D)},
	\end{split}
\end{equation}
 
In the same manner, we estimated the radial and orbital excited states of other singly, doubly, and triply bottom baryons for natural and unnatural parity states (see Tables \ref{tab:table12} to \ref{tab:table18}).

\subsection{Other states in the ($J,M^{2}$) plane}

So far we have calculated the masses of singly, doubly, and triply bottom baryons for natural and unnatural parity states by using the conventional formulas. After the successful implementation of this model, now in this section, we try to obtain the remaining other states in the ($J,M^{2}$) plane by using the same method. Since we have calculated $1^{2}P_{\frac{3}{2}}$ and $1^{4}P_{\frac{5}{2}}$ states earlier, now here we firstly calculate the other three $1P$ states i.e., $1^{2}P_{\frac{1}{2}}$, $1^{4}P_{\frac{1}{2}}$, and $1^{4}P_{\frac{3}{2}}$ by using the Eq. (\ref{eq:18}). For $\Omega_{b}^{-}$ baryon we put $i=u$, $j=s$, $q=b$, and $k=u$ in Eq. (\ref{eq:18}) we have,

\begin{eqnarray}
	\label{eq:33}
	\begin{split}
		& \left[(M_{\Sigma_{b}}+M_{\Omega_{b}})^{2}-4M^{2}_{\Xi_{b}^{'}}\right] \\
		& =
		\sqrt{(4M^{2}_{\Xi_{b}^{'}}-M^{2}_{\Sigma_{b}}-M^{2}_{\Omega_{b}})^{2}-4M^{2}_{\Sigma_{b}}M^{2}_{\Omega_{b}}}
	\end{split}
\end{eqnarray}

Due to the unavailability of experimental data, we have taken the masses of $\Sigma_{b}$ and
$\Xi_{b}^{'}$ baryons  from Ref. \cite{Ebert2011}. Hence after inserting $M_{\Sigma_{b}}$ and $M_{\Xi_{b}^{'}}$ for $1^{2}P_{\frac{1}{2}}$ state in Eq. (\ref{eq:33}), we get $M_{\Omega_{b}^{-}}$ = 6.365 GeV. Similarly we can obtain $M_{\Omega_{b}^{-}}$ = 6.359 GeV and $M_{\Omega_{b}^{-}}$ = 6.360 GeV for $1^{4}P_{\frac{1}{2}}$ and $1^{4}P_{\frac{3}{2}}$ states respectively. In the same manner we can calculate the masses for doubly and triply bottom baryons $\Xi_{bb}$, $\Omega_{bb}^{-}$ and $\Omega_{bbb}^{-}$ as we have done earlier. Inserting the masses of $\Lambda_{b}$, $\Sigma$, $\Xi$, and $\Xi_{b}$ baryons for $1^{2}P_{\frac{1}{2}}$ state taken from Ref. \cite{Ebert2011} and mass of $N$ baryon taken from Ref. \cite{Capstick1986} into Eq. (\ref{eq:20}), we get $M_{\Xi_{bb}}$ = 10.399 GeV for  $1^{2}P_{\frac{1}{2}}$ state. In this way all the remaining $1P$ states  for  $\Xi_{bb}$, $\Omega_{bb}^{-}$ and $\Omega_{bbb}^{-}$ baryons can be obtained in the same way as previously done.

Once we have calculated the other three $1P$ states- $1^{2}P_{\frac{1}{2}}$, $1^{4}P_{\frac{1}{2}}$, and $1^{4}P_{\frac{3}{2}}$, Regge slopes for these trajectories can be evaluated for singly, doubly, and triply bottom baryons using Eqs. (\ref{eq:3}), (\ref{eq:5}), and (\ref{eq:6}) with the same procedure as we have done above. The further excited state masses are calculated with the help  of the equation below,

\begin{equation}
	%\label{eq:29} 
	\nonumber
	M_{J+1} = \sqrt{M_{J}^{2}+\dfrac{1}{\alpha^{'}}} .
\end{equation}

\begin{table}%The best place to locate the table environment is directly after its first reference in text
	\caption{\label{tab:table19}
		Masses of other excited  states of $\Lambda_{b}^{0}$ baryon in the $(J,M^{2})$ plane. The numbers in the boldface are the theoretical values taken as the input \cite{Ebert2011} (in GeV).
	}
	\begin{ruledtabular}
		\begin{tabular}{llllllllll}
			\textit{$N^{2S+1}L_{J}$}& Present & PDG \cite{PDG} & \cite{B. Chen2015} &\cite{Mohammad} & \cite{Robert2008}& \cite{D.Jia} 
			\\	
			\colrule\noalign{\smallskip}
			
			$1^{2}P_{\frac{1}{2}}$ & \textbf{5.930} & 5.912 & 5.911 & 5.919 &5.939 &5.908 \\
			$1^{2}D_{\frac{3}{2}}$ &6.128 & 6.146 & 6.147 &6.199 &6.181 & 6.144 \\
			$1^{2}F_{\frac{5}{2}}$ & 6.320 & &6.346 &6.421 &6.206 &\\
			$1^{2}G_{\frac{7}{2}}$ & 6.506\\
			$1^{2}H_{\frac{9}{2}}$ & 6.687\\
			
		\end{tabular}
	\end{ruledtabular}
\end{table}

\begin{table}%The best place to locate the table environment is directly after its first reference in text
	\caption{\label{tab:table20}
		Masses of other excited  states of $\Sigma_{b}$ baryon in the $(J,M^{2})$ plane. The numbers in the boldface are the theoretical values taken as the input \cite{Ebert2011} (in GeV).
	}
	\begin{ruledtabular}
		
		\begin{tabular}{llllllllll}
			\textit{$N^{2S+1}L_{J}$}& Present &  \cite{Mohammad} &\cite{Rosner2015} &\cite{Yoshida2015} &   \\	
			\colrule\noalign{\smallskip}
			$1^{2}P_{\frac{1}{2}}$ & \textbf{6.101} &6.122&6.095 &6.127\\
			$1^{4}P_{\frac{1}{2}}$ & \textbf{6.095}& &6.087\\
			$1^{4}P_{\frac{3}{2}}$ & \textbf{6.087} & &6.096\\
			$1^{2}D_{\frac{3}{2}}$ & 6.293  &6.329\\
			$1^{4}D_{\frac{3}{2}}$ & 6.375&\\
			$1^{4}D_{\frac{5}{2}}$ & 6.346&\\
			$1^{2}F_{\frac{5}{2}}$ & 6.480 & 6.569\\
			$1^{4}F_{\frac{5}{2}}$ & 6.644\\
			$1^{4}F_{\frac{7}{2}}$ & 6.595\\
			$1^{2}G_{\frac{7}{2}}$ & 6.661\\
			$1^{4}G_{\frac{7}{2}}$ & 6.902\\
			$1^{4}G_{\frac{9}{2}}$ & 6.835\\
		\end{tabular}
	\end{ruledtabular}
\end{table}

\begin{table}%The best place to locate the table environment is directly after its first reference in text
	\caption{\label{tab:table21}
		Masses of other excited states of $\Xi_{b}$ baryon in the $(J,M^{2})$ plane. The numbers in the boldface are the theoretical values taken as the input \cite{Ebert2011} (in GeV).
	}
	\begin{ruledtabular}
		
		\begin{tabular}{llllllllll}
			\textit{$N^{2S+1}L_{J}$}& Present & \cite{B. Chen2015} & \cite{Robert2008}  &\cite{Z. Wang 2010.11}    \\	
			\colrule\noalign{\smallskip}
			$1^{2}P_{\frac{1}{2}}$ & \textbf{6.120}&6.097 &6.090 &\\
			$1^{4}P_{\frac{1}{2}}$ &\textbf{6.227}& & & 6.140\\
			$1^{4}P_{\frac{3}{2}}$ & \textbf{6.224}\\
			$1^{2}D_{\frac{3}{2}}$ & 6.316 &6.344\\
			$1^{4}D_{\frac{3}{2}}$ & 6.508\\
			$1^{4}D_{\frac{5}{2}}$ & 6.484\\
			$1^{2}F_{\frac{5}{2}}$ & 6.506 &6.555\\
			$1^{4}F_{\frac{5}{2}}$ & 6.777\\
			$1^{4}F_{\frac{7}{2}}$ & 6.734\\
			$1^{2}G_{\frac{7}{2}}$ & 6.690&6.743\\
			$1^{4}G_{\frac{7}{2}}$ & 7.036\\
			$1^{4}G_{\frac{9}{2}}$ & 6.974\\
		\end{tabular}
	\end{ruledtabular}
\end{table}

\begin{table}%The best place to locate the table environment is directly after its first

	\caption{\label{tab:table22}
		Masses of other excited states of $\Xi_{b}^{'}$ baryon in the $(J,M^{2})$ plane. The numbers in the boldface are the theoretical values taken as the input \cite{Ebert2011} (in GeV).
	}
	\begin{ruledtabular}
		\begin{tabular}{llllllllll}
			\textit{$N^{2S+1}L_{J}$}& Present & \cite{Robert2008} &\cite{Wei.2017}
			\\	
			\colrule\noalign{\smallskip}
			
			$1^{2}P_{\frac{1}{2}}$ & \textbf{6.233} &6.305\\
			$1^{2}D_{\frac{3}{2}}$ & 6.425 \\
			$1^{2}F_{\frac{5}{2}}$ & 6.612 &\\
			$1^{2}G_{\frac{7}{2}}$ & 6.794\\
			$1^{2}H_{\frac{9}{2}}$ & 6.971\\
			
		\end{tabular}
	\end{ruledtabular}
\end{table}

\begin{table}%The best place to locate the table environment is directly after its first reference in text
	\caption{\label{tab:table23}
		Masses of other excited states of $\Omega_{b}^{-}$ baryon in the $(J,M^{2})$ plane.
	}
	\begin{ruledtabular}
		
		\begin{tabular}{llllllllll}
			\textit{$N^{2S+1}L_{J}$}& Present & PDG \cite{PDG} & \cite{Mohammad}& \cite{H. X. Chen2020} &\cite{Z. G. Wang2020} & \\	
			\colrule\noalign{\smallskip}
			$1^{2}P_{\frac{1}{2}}$ & 6.365 & &6.342&6.340 &6.330  \\
			$1^{4}P_{\frac{1}{2}}$ &6.359 & & &6.340 \\
			$1^{4}P_{\frac{3}{2}}$ &6.360&6.350 & &6.350 &6.350 \\
			$1^{2}D_{\frac{3}{2}}$ &6.557& &6.545 \\
			$1^{4}D_{\frac{3}{2}}$ &6.640 \\
			$1^{4}D_{\frac{5}{2}}$ &6.620 \\
			$1^{2}F_{\frac{5}{2}}$ &6.744 & &6.777 \\
			$1^{4}F_{\frac{5}{2}}$ &6.909 \\
			$1^{4}F_{\frac{7}{2}}$ &6.870 \\
			$1^{2}G_{\frac{7}{2}}$ &6.926\\
			$1^{4}G_{\frac{7}{2}}$ &7.168 \\
			$1^{4}G_{\frac{9}{2}}$ &7.111 \\
		\end{tabular}
	\end{ruledtabular}
\end{table}

\begin{table}%The best place to locate the table environment is directly after its first reference in text
	\caption{\label{tab:table24}
		Masses of other excited states of $\Xi_{bb}$ baryon in the $(J,M^{2})$ plane. 
	}
	\begin{ruledtabular}
		
		\begin{tabular}{llllllllll}
			\textit{$N^{2S+1}L_{J}$}& Present &\cite{S. Gershtein2000} &\cite{Yoshida2015} &\cite{Valcarce2008} &\cite{Ebert2005.2.8} & \\	
			\colrule\noalign{\smallskip}
			$1^{2}P_{\frac{1}{2}}$ &10.399 &10.310 &10.476 &10.406 &10.368\\
			$1^{4}P_{\frac{1}{2}}$ &10.551 &10.541\\
			$1^{4}P_{\frac{3}{2}}$ &10.557 &10.567\\
			$1^{2}D_{\frac{3}{2}}$ &10.556 \\
			$1^{4}D_{\frac{3}{2}}$ &10.810 \\
			$1^{4}D_{\frac{5}{2}}$ &10.795 \\
			$1^{2}F_{\frac{5}{2}}$ &10.711 \\
			$1^{4}F_{\frac{5}{2}}$ &11.062 \\
			$1^{4}F_{\frac{7}{2}}$ &11.028 \\
			$1^{2}G_{\frac{7}{2}}$ &10.864 \\
			$1^{4}G_{\frac{7}{2}}$ &11.309 \\
			$1^{4}G_{\frac{9}{2}}$ &11.256 \\
		\end{tabular}
	\end{ruledtabular}
\end{table}

\begin{table}%The best place to locate the table environment is directly after its first reference in text
	\caption{\label{tab:table25}
		Masses of other excited states of $\Omega_{bb}^{-}$ baryon in the $(J,M^{2})$ plane.
	}
	\begin{ruledtabular}
		
		\begin{tabular}{llllllllll}
			\textit{$N^{2S+1}L_{J}$}& Present &\cite{Yoshida2015} &\cite{Ebert2005.2.8} &\cite{Valcarce2008} &\cite{T.M. Aliev2014}  \\	
			\colrule\noalign{\smallskip}
			$1^{2}P_{\frac{1}{2}}$ & 10.546 &10.607 &10.532 &10.519\\
			$1^{4}P_{\frac{1}{2}}$ & 10.678\\
			$1^{4}P_{\frac{3}{2}}$ & 10.688& & & &10.513\\
			$1^{2}D_{\frac{3}{2}}$ & 10.706\\
			$1^{4}D_{\frac{3}{2}}$ & 10.952\\
			$1^{4}D_{\frac{5}{2}}$ & 10.940\\
			$1^{2}F_{\frac{5}{2}}$ & 10.863\\
			$1^{4}F_{\frac{5}{2}}$ & 11.219\\
			$1^{4}F_{\frac{7}{2}}$ & 11.187\\
			$1^{2}G_{\frac{7}{2}}$ & 11.018\\
			$1^{4}G_{\frac{7}{2}}$ & 11.480\\
			$1^{4}G_{\frac{9}{2}}$ & 11.428\\
		\end{tabular}
	\end{ruledtabular}
\end{table}

\begin{table}%The best place to locate the table environment is directly after its first reference in text
	\caption{\label{tab:table26}
		Masses of other excited states of $\Omega_{bbb}^{-}$ baryon in the $(J,M^{2})$ plane. 
	}
	\begin{ruledtabular}
		
		\begin{tabular}{llllllllll}
			\textit{$N^{2S+1}L_{J}$}& Present &\cite{Robert2008} &\cite{Wei.2017} & \cite{Z.G. Wang2012}&\cite{T.M. Aliev2014}  \\	
			\colrule\noalign{\smallskip}
			
			$1^{4}P_{\frac{1}{2}}$ & 14.999 &14.975 &\\
			$1^{4}P_{\frac{3}{2}}$ & 15.017&14.976 &15.055 &14.950 &14.900\\
			
			$1^{4}D_{\frac{3}{2}}$ & 15.267\\
			$1^{4}D_{\frac{5}{2}}$ & 15.263&15.101\\
			
			$1^{4}F_{\frac{5}{2}}$ & 15.530\\
			$1^{4}F_{\frac{7}{2}}$ & 15.506\\
			
			$1^{4}G_{\frac{7}{2}}$ & 15.789\\
			$1^{4}G_{\frac{9}{2}}$ & 15.745\\
		\end{tabular}
	\end{ruledtabular}
\end{table}

\begin{table*}%The best place to locate the table environment is directly after its first reference in text
	\caption{\label{tab:table27}
		Masses of excited states of $\Omega_{b}^{-}$ baryon in the $(J,M^{2})$ plane (in GeV).
	}
	\begin{ruledtabular}
		
		\begin{tabular}{llllllllllllll}
			\textit{$N^{2S+1}L_{J}$} & Present & PDG \cite{PDG} &\cite{Thakkar2017} &\cite{Ebert2011} &\cite{Wei.2017} &\cite{Yoshida2015} & \cite{Mohammad}& \cite{H. X. Chen2020} &\cite{Z. G. Wang2020} \\	
			\colrule\noalign{\smallskip}
			$1^{2}S_{\frac{1}{2}}$ & 6.054 &6.046 &6.048 &6.064 &6.048 &6.076 &6.098\\
			$1^{4}S_{\frac{3}{2}}$ & 6.074 & &6.086 &6.088 &6.069 &6.094  \\
			
			$1^{2}P_{\frac{1}{2}}$ & 6.365& & & & & & 6.342&6.340 &6.330  \\
			$1^{2}P_{\frac{3}{2}}$ & 6.348 &6.340 &6.328 &6.340 &6.325 &6.336 \\
		    $1^{4}P_{\frac{1}{2}}$ &6.359 & & & && & &6.340 \\
			$1^{4}P_{\frac{3}{2}}$ &6.360&6.350 & & & & & &6.350 &6.350 \\
			$1^{4}P_{\frac{5}{2}}$ & 6.362 & &6.320 &6.334 &6.345 &6.345 \\
			
			$1^{2}D_{\frac{3}{2}}$ &6.557& & & & & &6.545 \\
			$1^{2}D_{\frac{5}{2}}$ & 6.629 & &6.567 &6.529 &6.590 &6.561\\
			$1^{4}D_{\frac{3}{2}}$ &6.640 \\
			$1^{4}D_{\frac{5}{2}}$ &6.620 \\
			$1^{4}D_{\frac{7}{2}}$ & 6.638 & &6.553 &6.517 &6.609\\
			
			$1^{2}F_{\frac{5}{2}}$ &6.744 & & & & & &6.777 \\
			$1^{2}F_{\frac{7}{2}}$ &6.899 & &6.800 &6.736 &6.844\\
			$1^{4}F_{\frac{5}{2}}$ &6.909 \\
			$1^{4}F_{\frac{7}{2}}$ &6.870 \\
			$1^{4}F_{\frac{9}{2}}$ & 6.903 & &6.780 &6.713 &6.863 \\
			
			$1^{2}G_{\frac{7}{2}}$ &6.926\\
			$1^{2}G_{\frac{9}{2}}$ & 7.159 & & &6.915 &7.090\\
			$1^{4}G_{\frac{7}{2}}$ &7.168 \\
			$1^{4}G_{\frac{9}{2}}$ &7.111 \\
			$1^{4}G_{\frac{11}{2}}$& 7.158 & & &6.884 &7.108 \\
			
			$1^{2}H_{\frac{11}{2}}$& 7.409 \\
			$1^{4}H_{\frac{13}{2}}$& 7.404 \\	
						\end{tabular}
			 The excited state masses of $\Omega_{b}^{-}$ baryon including natural, unnatural, and other states. \\
				\end{ruledtabular}
				%\hline
\end{table*}

Tables \ref{tab:table19} - \ref{tab:table23} shows our calculated results for the remaining other states for singly bottom baryons. We compared our estimated masses with other theoretical studies and our results are in accordance with them. Similarly tables \ref{tab:table24} - \ref{tab:table26} shows our predicted masses for doubly and triply bottom baryons. Very few results are available from previous theoretical studies and they are consistent with our estimated masses. 

\section{\label{sec:level2}Results and discussion}
In the present work, under the methodology of Regge phenomenology, we have obtained the ground state masses of unseen $\Omega_{b}^{-}$, $\Xi_{bb}^{0}$, $\Xi_{bb}^{-}$, $\Omega_{bb}^{-}$, and $\Omega_{bbb}^{*-}$ baryons. Regge slopes of the singly, doubly, and triply bottom baryon trajectories were calculated in the $(J,M^{2})$ plane. With the aid of these Regge slopes, the masses of the orbitally excited bottom baryons were estimated for both natural and unnatural parity states. After that, the Regge slopes and intercepts were extracted for each Regge line in the $(n,M^{2})$ plane, and with the help of these parameters mass spectra of singly, doubly, and triply bottom baryons were obtained successfully.

\begin{enumerate}
\item\textbf{$\Lambda_{b}^{0}$ baryon}: Experimentally, there are quite number of excited $\Lambda_{b}^{0}$ states reported recently; $\Lambda_{b}(5912)^{0}$, $\Lambda_{b}(5920)^{0}$, $\Lambda_{b}(6070)^{0}$, $\Lambda_{b}(6146)^{0}$, and $\Lambda_{b}(6152)^{0}$. Our calculated masses for $\Lambda_{b}^{0}$ baryon in the $(J,M^{2})$ plane are shown in Table \ref{tab:table4} for natural parity states. We observed that the experimentally found state $\Lambda_{b}(5920)^{0}$ having mass 5.920 GeV is close to our predicted mass 5.924 GeV with a slight mass difference of 4 MeV. So we confirmed this state as $1P$ state with $J^{P}=\frac{3}{2}^{-}$. Also, the $\Lambda_{b}(6152)^{0}$ having mass 6.152 GeV is somewhat lower than our estimated mass 6.213 GeV with a mass difference of 61 MeV. So this state is confirmed to be $1D$ state with $J^{P}=\frac{5}{2}^{+}$. Other than these two, one more state  $\Lambda_{b}(6146)^{0}$ having mass 6.146 GeV is close to our calculated mass 6.128 GeV with a mass difference of 18 MeV. Hence we confirmed this state as $1D$ state with $J^{P}=\frac{3}{2}^{+}$ (see Table \ref{tab:table19}). Further we compared excited state masses with other theoretical predictions \cite{Thakkar2017,Ebert2011,Ebert2005.2.8,Wei.2017} and our results are in agreement with them. Similarly, we compared our results evaluated in the ($n,M^{2}$) plane and they are close to other theoretical and phenomenological studies \cite{Ebert2011,Yoshida2015} as shown in Table \ref{tab:table12}.     
	
\item \textbf{$\Sigma_{b}^{\pm}$ baryons}: Table \ref{tab:table5} shows our calculated results for $\Sigma_{b}^{\pm}$ baryons for natural and unnatural parity states in the $(J,M^{2})$ plane. Experimentally, two resonances $\Sigma_{b}(6097)^{+}$ and $\Sigma_{b}(6097)^{-}$ were discovered recently having masses 6.096 GeV and 6.098 GeV respectively, which are very close to our estimated masses 6.105 GeV ($\Sigma_{b}^{+}$) and 6.110 GeV ($\Sigma_{b}^{-}$) with a mass difference of 9-12 MeV. Hence, $\Sigma_{b}(6097)^{\pm}$ states are identified as $1P$-states with $J^{P}=\frac{3}{2}^{-}$ for S = 1/2. Our estimated results are also in accordance with other theoretical predictions \cite{Ebert2011,Wei.2017,Yoshida2015}. In the same manner we compared our calculated radial and orbital excited state masses evaluated in the ($n,M^{2}$) plane for both natural and unnatural parity states (see Table \ref{tab:table13}), and our results are consistent with other theoretical predictions.
	
\item \textbf{$\Xi_{b}^{0,-}$, $\Xi_{b}^{'-}$ baryons}: Orbitally excited state masses for $\Xi_{b}^{0,-}$ and $\Xi_{b}^{'-}$ baryons are shown in Table \ref{tab:table6} and \ref{tab:table7} respectively. We compared our calculated results with other theoretical predictions and experimental results. Experimentally found new state $\Xi_{b}(6227)^{-}$ having mass 6.227 GeV is close to our predicted mass 6.229 GeV (shown in Table \ref{tab:table7}). This state can be a good candidate of $P$-wave and identified with spin parity $J^{P}=\frac{3}{2}^{-}$ for S=1/2. In the $(n,M^{2})$ plane our calculated results for $\Xi_{b}^{0,-}$, and $\Xi_{b}^{'-}$ baryons are consistent with the predictions of \cite{Ebert2011,Wei.2017,Robert2008,H. Garcilazo2007} (see Tables \ref{tab:table14} and \ref{tab:tablecas}).
	
\item \textbf{$\Omega_{b}^{-}$ baryon}: Our estimated ground and excited state masses of $\Omega_{b}^{-}$ baryon in the $(J,M^{2})$ and $(n,M^{2})$ plane are shown in Table \ref{tab:table8} and \ref{tab:table15} respectively. Experimentally, only the ground state $\Omega_{b}^{-}$ is observed with spin parity $J^{P}=\frac{1}{2}^{+}$, $\Omega_{b}^{*-}$ with $J^{P}=\frac{3}{2}^{+}$ is still unseen. Our calculated ground-state $\Omega_{b}^{-}$ having a mass of 6.054 GeV is very close to the experimental data \cite{PDG} with mass difference of 8 MeV. Also, for $1^{4}S_{\frac{3}{2}}$ state our predicted mass is in accordance with the results of Ref. \cite{Thakkar2017,Ebert2011,Wei.2017,Yoshida2015}. After that, we compared our calculated orbital and radial excited states for natural and unnatural parity states in the ($J,M^{2}$) and ($n,M^{2}$) planes with other theoretical results. Our predicted masses are consistent with them. The experimentally observed state $\Omega_{b}(6340)^{-}$ having a mass of 6.340 GeV is close to our estimated mass of 6.348 GeV with a mass difference of 8 MeV. So, we assigned $\Omega_{b}(6340)^{-}$ as $1P$ state with $J^{P} = 3/2^{-}$ for S = 1/2. Also, one more state $\Omega_{b}(6350)^{-}$ is found near to our calculated mass 6.360 GeV, hence wecan say that this state may belong to $1P$ state with $J^{P} = 3/2^{-}$ for S = 3/2 (see table \ref{tab:table23}).  In the present work, other two states $\Omega_{b}(6316)^{-}$, and $\Omega_{b}(6330)^{-}$ are not identified.
	
\item \textbf{$\Xi_{bb}^{0,-}$ baryons}:  Table \ref{tab:table9} shows our calculated orbitally excited state masses for $\Xi_{bb}^{0}$ and $\Xi_{bb}^{-}$ baryons for both natural and unnatural parities in the ($J,M^{2}$) plane.  Our estimated ground state ($1^{2}S_{\frac{1}{2}}$) masses are in good agreement with the prediction of Ref. \cite{Wei.2017} with a mass difference of 26-31 MeV. Similarly, for $1^{4}S_{\frac{3}{2}}$ state, our predicted masses are close to the results of Refs. \cite{Wei.2017,B. Eakins2012} with a slight mass difference of 9-17 MeV. Also, the mass splitting $M_{\Xi_{bb}^{0}}-M_{\Xi_{bb}^{*0}}$ = 105 MeV is little big, however it is cose to Refs. \cite{Wei.2017,Z. Ghalenovi2011,Ghalenovi2014}. Further, the excited state masses are also in accordance with the predictions of other theoretical studies \cite{Wei.2017,Robert2008,B. Eakins2012}. In the same manner, we compared our evaluated excited state masses in ($n,M^{2}$) plane for natural and unnatural parities with other theoretical predictions  and phenomenological studies as shown in Table \ref{tab:table16}. Our results are consistent with the predictions of \cite{Ebert2005.2.8,Yoshida2015,Robert2008,B. Eakins2012,F.Giannuzzi2009}.
	
\item \textbf{$\Omega_{bb}^{-}$ baryon}: Our calculated ground and excited state masses of $\Omega_{bb}^{-}$ baryon for natural and unnatural parities in the ($J,M^{2}$) plane are shown in Table  \ref{tab:table10}. The ground state ($1^{4}S_{\frac{1}{2}}$) mass is close to the masses of Refs. \cite{Wei.2017,Ebert2005.2.8} and somewhat lower than the results of \cite{Yoshida2015,Robert2008,Zalak.bb2016}. For $1^{4}S_{\frac{3}{2}}$ state, our estimated mass is very well fitted with the results of the Refs. \cite{Wei.2017,Yoshida2015,Zalak.bb2016}. Here also the mass splitting $M_{\Omega_{bb}^{-}}-M_{\Omega_{bb}^{*-}}$ = 99 MeV, is reasonable with Refs. \cite{Wei.2017,Z. Ghalenovi2011,Ghalenovi2014}. For excited states, our predicted masses are reasonably close to the results of \cite{Wei.2017} and they are also in accordance with the other theoretical studies. Similarly, Table \ref{tab:table17} shows our estimated radial and orbital excited states in the ($n,M^{2}$) plane. Our results are consistent with the predictions of other theoretical approaches.

\item \textbf{$\Omega_{bbb}^{*-}$ baryon}: For triply bottom baryon, we obtained the ground state and excited state masses in both the ($J,M^{2}$) and ($n,M^{2}$) planes as shown in Table \ref{tab:table11} and \ref{tab:table18} respectively. The ground state ($1^{4}S_{\frac{3}{2}}$) mass for $\Omega_{bbb}^{*-}$ vary in the range 14.360-15.130 GeV predicted in other theoretical approaches and phenomenological studies (see Table \ref{tab:table2}). Our estimated ground-state mass shows a few-MeV difference with the results of \cite{Wei.2017,Z. Wang 2010.11,Robert2008}. For excited states, very limited results are available from previous theoretical studies and our estimated masses are in accordance with them. 
\end{enumerate}

\section{Conclusion}

Here, our aim is satisfied for the determination of spin-parity of experimentally observed unknown states: $\Sigma_{b}(6097)^{+}$, $\Sigma_{b}(6097)^{-}$, $\Xi_{b}(6227)^{-}$,  $\Omega_{b}(6340)^{-}$, and $\Omega_{b}(6350)^{-}$. Also, we confirmed the spin-parity of $\Lambda_{b}(5920)^{0}$, $\Lambda_{b}(6146)^{0}$, and $\Lambda_{b}(6152)^{0}$. This model is successful for the study of singly, doubly, and triply bottom baryons. The masses of the orbitally excited bottom baryons were estimated for both natural and unnatural parity states in the $(J,M^{2})$ and $(n,M^{2})$ plane and also the other states in the $(J,M^{2})$ plane. Our predictions will help future experimental studies at LHCb, CMS, and Belle II to identify these baryonic states. The reliability of this model is very much dependent on the availability of experimental data that we have taken as input.    

\begin{acknowledgments}
	J. O. was inspired by the work of  Ke-Wei Wei, Bing Chen, Xin-Heng Guo, De-Min Li, Bing Ma, Yu-Xiao Li, Qian-Kai Yao, and Hong Yu on Regge phenomenology and would like to thank them for their valuable contributions to this field.
\end{acknowledgments}

\end{document}